\documentclass[3p,preview,authoryear]{elsarticle}
 
\usepackage{booktabs}           
\usepackage{tikz}               
\usepackage{mathtools}               

\usepackage{lineno,hyperref}
\usepackage[symbol]{footmisc}

\usepackage{sectsty}
\usepackage{fullpage}
\usepackage{graphicx}
\usepackage{caption}
\usepackage{epstopdf}
\usepackage{epsfig}
\usepackage{pdfpages}
\usepackage{float}
\usepackage{amsmath,bm}
\usepackage{textcomp}
\usepackage{upgreek}
\usepackage{multirow}
\usepackage{textcomp}
\usepackage{accents}
\usepackage{amsfonts}
\usepackage{lscape}
\usepackage{comment}

\modulolinenumbers[1]

\journal{arXiv}

\begin{document}

\begin{frontmatter}

\title{Effect of restrained versus free drying on hygro-expansion of hardwood and softwood fibers and paper handsheet}

\author[1]{N.H. Vonk} \author[1]{R.H.J. Peerlings} \author[1]{M.G.D. Geers} \author[1]{J.P.M. Hoefnagels*} 

\cortext[mycorrespondingauthor]{J.P.M. Hoefnagels}
\ead{J.P.M.Hoefnagels@tue.nl}

\address[1]{Department of Mechanical Engineering, Eindhoven University of Technology, Eindhoven, The Netherlands}

\begin{abstract}
Earlier works in literature on the hygro-expansion of paper state that the larger hygro-expansivity of freely compared to restrained dried handsheets is due to structural differences between the fibers inside the handsheet. To unravel this hypothesis, first, the hygro-expansion of freely and restrained dried, hardwood and softwood handsheets has been characterized. Subsequently, the transient full-field hygro-expansion (longitudinal, transverse, and shear strain) of fibers extracted from these handsheets was measured using global digital height correlation, from which the micro-fibril angle was deduced. The hygro-expansivity of each individual fiber was tested before and after a wetting period, during which the fiber’s moisture content is maximized, to analyze if a restrained dried fiber can “transform” into a freely dried fiber. It was found that the longitudinal hygro-expansion of the freely dried fibers is significantly larger than the restrained dried fibers, consistent with the sheet-scale differences. The difference in micro-fibril angle between the freely and restrained dried fibers is a possible explanation for this difference, but merely for the hardwood fibers, which are able to “transform” to freely dried fibers after being soaked in water. In contrast, this "transformation" does not happen in softwood fibers, even after full immersion in water for a day. Various mechanisms have been studied to explain the observations on freely and restrained dried hardwood and softwood, fiber and handsheets including analysis of the fibers’ lumen and cross-sectional shape. The presented results and discussion deepens the understanding of the differences between freely and restrained dried handsheets. \\ 
\end{abstract}


\end{frontmatter}

\newpage
\section{Introduction}
Paper is a composite material consisting of natural fibers, e.g., hardwood or softwood. The extreme deterioration of paper's mechanical and geometrical properties to moisture poses one of the key problems in its exploitation, possibly resulting in non-usable end products. Therefore, being able to control the moisture-induced dimensional stability of paper remains a great challenge. This is especially true for printing applications, during which paper is subjected to multiple drying and wetting cycles while being, respectively, constrained and unconstrained. Failure to control the paper sheet deformation can result in out-of-plane deformations, including curling, cockling, fluting and waviness \citep{kulachenko2005tension, de2017moisture}, which are driven by processes occurring in the fibrous microstructure, down to the single fiber level. The severity of these out-of-plane deformations are linked to the magnitude of the moisture-induced dimensional change (hygro-expansion) of the paper sheet.\\ \indent
Before addressing the hygro-expansivity of paper, first the paper fiber structure is briefly discussed. Paper fibers are open tube-like structures with a tube wall that consists of four layers (P, S1, S2, S3) made of cellulose, hemi-celluloses and lignin \citep{liitia2000solid, hubbe2014prospects}. The lumen, i.e. the tube opening, may be collapsed according to the fiber's processing conditions, i.e. pulping, refining, or paper drying pressure \citep{hubbe2014prospects, he2003behavior, hubbe2007happens}. The S2 layer constitutes approximately 80\% of the fiber wall \citep{courchene2006cellulose, cristian2006ultrastructural}, and is likely to dominate the fiber's hygro-mechanical properties. Each layer is built of long parallel cellulose strands, called micro-fibrils, which alternate between crystalline (40$-$60 vol\%) and so-called "dislocated regions" \citep{agarwal2013estimation}. These micro-fibrils are helically oriented in each sub-layer, and are stiffer than the hemi-celluloses and lignin counterparts that bind the micro-fibrils into the layer \citep{hubbe2014prospects, salmen1982temperature, haslach2000moisture, fahlen2005pore, berglund201112}. The angle between the micro-fibrils' axis and fiber's longitudinal axis is called the micro-fibril angle (MFA), which is different for each layer and for each wood type. The MFA (of the S2 layer) of hardwood, such as Eucalyptus, is between 0 and 11\textsuperscript{o}, whereas softwood, such as a mixture of Spruce and Pine, typically has a (much) higher MFA ranging from 8 to 39\textsuperscript{o} \citep{french2000effect, barnett2004cellulose, cown2004wood, donaldson2008microfibril}. The significantly larger variation in MFA observed for softwood fibers is due to the large difference in MFA between early- and latewood fibers \citep{anagnost2002variation, jordan2005multilevel}. Since the MFA is low for most fibers, the fiber exhibits strong anisotropic hygro-mechanical properties \citep{retulainen1998papermaking}, i.e., a longitudinal to transverse stiffness ratio of 6$-$11 as found in experiments \citep{czibula2021transverse} and used in models \citep{magnusson2013numerical, brandberg2020role}. Furthermore, the fibers exhibit a longitudinal to transverse hygro-expansion ratio of 20$-$40 \citep{wahlstrom2009development, joffre2016method, vonk2021full}. The key challenge now remains how these strongly anisotropic fiber properties translate to the network-scale hygro-expansion, where, e.g., in the bonded areas a competition between the fibers arises. \\ \indent
During paper sheet wetting, the longitudinal fiber hygro-expansion directly contributes to the sheet-scale hygro-expansion through the extension of the freestanding segments. The much larger transverse strain only contributes indirectly to the sheet by transmission through the inter-fiber bonds to the longitudinal extension of the adhered fibers \citep{brandberg2020role}. Theoretical and numerical models have shown that the transverse fiber strain contribution to the sheet scale is relatively weak, i.e. the longitudinal fiber hygro-expansivity dominates \citep{brandberg2020role, uesaka1994general, motamedian2019simulating}. However, detailed mechanistic small-scale experiments of the transverse and longitudinal fiber hygro-expansion are lacking, whereas these experiments are essential for tailoring the paper structure to improve its dimensional stability. \cite{lindner2018factors} wrote an outstanding review on factors affecting the dimensional stability of paper, in which, the author indicated that increasing the level of restrained drying strongly reduces the hygro-expansivity of paper. However, the driving mechanisms behind these phenomena have not been unraveled, which constitutes the main objective of this work. \\ \indent
Two paper types are usually considered when investigating the effect of a drying restraint; machine paper with fibers oriented in the machine direction (MD) making the hygro-mechanical behavior anisotropic, and handsheets which have a random fiber orientation and isotropic behavior. For commercial machine paper making, tension is applied in MD \citep{nanko1995mechanisms, makela2009effect}, inducing a restrained MD, but, depending on the draw, also a restrained CD. Regarding handsheet paper making, restrained dried (RD) handsheets are formed by applying a constant out-of-plane pressure to wet webs, while freely dried (FD) handsheets are typically formed by drying the wet web in between PTFE meshes \citep{uesakaQi1994hygroexpansivity, larsson2008influence, fellers2007interaction, urstoger2020microstructure}. After RD paper formation, dried-in strain is stored inside the paper, which upon introduction of moisture is released, resulting in an irreversible shrinkage \citep{larsson2008influence, uesaka1992characterization}. Interestingly, in \citep{vonk2021full, vonk2020robust}, it was found that this release of dried-in strain actually emerges from the single fiber level. Regarding the mechanical properties, lab-made machine paper which was restrained in both CD and MD exhibits a higher specific modulus in both directions compared to FD machine paper \citep{uesakaQi1994hygroexpansivity}, and the same is found for handsheets, where RD handsheets exhibit a lower strain to failure and larger Young's modulus than FD \citep{makela2009effect, urstoger2020microstructure, kouko2014influence}. Regarding hygro-expansivity, machine paper dried under restrained in CD and MD exhibits a smaller hygro-expansivity than FD machine paper \citep{uesakaQi1994hygroexpansivity}. Similarly, RD handsheets exhibit a significantly smaller hygro-expansivity than FD handsheets \citep{larsson2008influence, fellers2007interaction, urstoger2020microstructure, uesaka1992characterization, salmen1987development, salmen1987implications, nanri1993dimensional}. Hence, RD increases the Young's modulus and dimensional stability of machine paper (in MD and CD) and handsheets compared to FD. The open scientific question that remains is what drives these differences. This work focuses on unraveling the mechanics governing the hygro-expansivity. \\ \indent
\cite{uesakaQi1994hygroexpansivity} stated that the fibers constituting the inter-fiber bonds inside FD handsheets are more wrapped around each other than their RD handsheet counterparts, due to the absence of an external applied pressure for FD. Consequently, the increased bonded surface would result in a larger transverse strain transfer in the bonded region. This theory is adopted in a few other works \citep{nanko1995mechanisms, larsson2008influence, fellers2007interaction}. However, recent work by \cite{urstoger2020microstructure} demonstrated, using 3D X-ray computed tomography characterization of the inter-fiber bond geometries, that the difference in wrap around angle inside FD and RD handsheets is negligible and not significant enough to explain the large hygro-expansivity magnitudes. \\ \indent
In older works, \cite{van1961some} convincingly argued, based on machine paper experiments, that the difference in mechanical properties between the CD and MD can not solely be explained by the fiber orientation, but are expected to be caused by differences in the drying procedure (i.e. constrained during drying). \cite{jentzen1964effect} showed strong evidence that fibers dried under a constant tensile stress reveal a higher Young's modulus than fibers dried under no stress. The author attributed this to the lower MFA of fibers dried under stress. This would mean that changes in the fiber's structure cause the difference in mechanical properties. Interestingly, \cite{meylan1972influence} and \cite{yamamoto2001model} showed, for wood fibers, that the magnitude of longitudinal fiber shrinkage decreases for decreasing MFA, which, combined with the work of \cite{nanko1995mechanisms} showing that fibers dried in MD exhibit a lower longitudinal shrinkage than fibers in CD, might suggest that the fibers' structural differences affect the fibers' hygro-expansivity and consequently explains the hygro-expansivity difference of FD and RD paper sheets. Additionally, \cite{salmen1987development} proposed a theory on how dried-in strain is stored inside the amorphous hemi-cellulose and "dislocated cellulose regions" of the fiber and how this affects the hygro-expansivity of the fiber itself. In order to test if structural changes in the fibers can explain the hygro-expansivity differences, one should test single fibers extracted from FD and RD paper and relate the findings at the micro- (fiber) and macro-scale (sheet). \\ \indent
Accordingly, in this work the hygro-expansivity of fibers isolated from either RD or FD handsheets, made of either hardwood or softwood, are tested using a recently developed full-field hygro-expansivity method \citep{vonk2020robust}, and compared to the sheet-scale hygro-expansivity. Furthermore, all fibers are loaded to maximum moisture content to determine the release of dried-in strain. Surprisingly, not all fibers are able to release their dried-in strain, and hence additional experiments considering the fiber structure are performed. This approach enables (i) investigation of the mechanisms driving the hygro-expansivity difference, (ii) direct fiber to sheet hygro-expansivity comparison, allowing to distinguish the longitudinal and transverse fiber hygro-expansion contribution to the sheet scale. In this work, first the handsheet and fiber preparation is elaborated, followed by the specific testing procedure of the handsheets and fibers. Then the results are given and discussed in which some additional experiments are elaborated, and, finally, the main conclusions are given\footnote[1]{Note that a few preliminary experiments of this work have already been reported in \citep{vonk2023frc}, including the hygro-expansion of fibers extracted from hardwood handsheets, of which the main purpose was to optimize and improve the method's resolution and procedure for the current work, which reveals novel insights.}. 

\section{Materials and Methods}

\subsection{Preparation of the handsheets}
FD and RD handsheets with an average weight of \textsuperscript{$\sim$}60 g/m\textsuperscript{2} produced from bleached hardwood (HW) (Eucalyptus) and softwood (SW) (mixture of Spruce and Pine) kraft pulp (kappa $<$2), using the “Rapid Kötchen” device, were kindly provided by \textit{Mondi Group, Austria}. The method proposed by \cite{larsson2008influence} was followed; the FD handsheets were once pressed for 1 minute with a temperature ($T$) of 93\textsuperscript{o}C and pressure of 95 kPa and afterwards freely dried in a PTFE mesh drying frame consisting of stabilizing bars. The RD handsheets were pressed for 10 minutes with a $T$ of 93\textsuperscript{o}C and pressure of 95 kPa. Note that the used HW and SW pulps (stored at RH = 50\% and $T$ = 23\textsuperscript{o}C) are the same as used for the single fiber hygro-expansion experiments conducted in \citep{vonk2021full}, thus enabling direct comparison of results. Each handsheet was subsequently cut into 6$\times$6 cm\textsuperscript{2} sheets and a random speckle pattern was applied using charcoal sticks as shown in Figure \ref{fig:sheet_method}, instead of using spray paint, of which the solvent could affect the paper structure. The speckle pattern is required for a Global Digital Image Correlation (GDIC) algorithm as proposed by \cite{neggers2016image} to track the displacement field, which enables identification of the in-plane hygro-expansion.

\subsection{Preparation of the single fibers}
\begin{figure}[]
	\centering
	\includegraphics[width=0.48\textwidth,trim=6 4 6 4,clip]{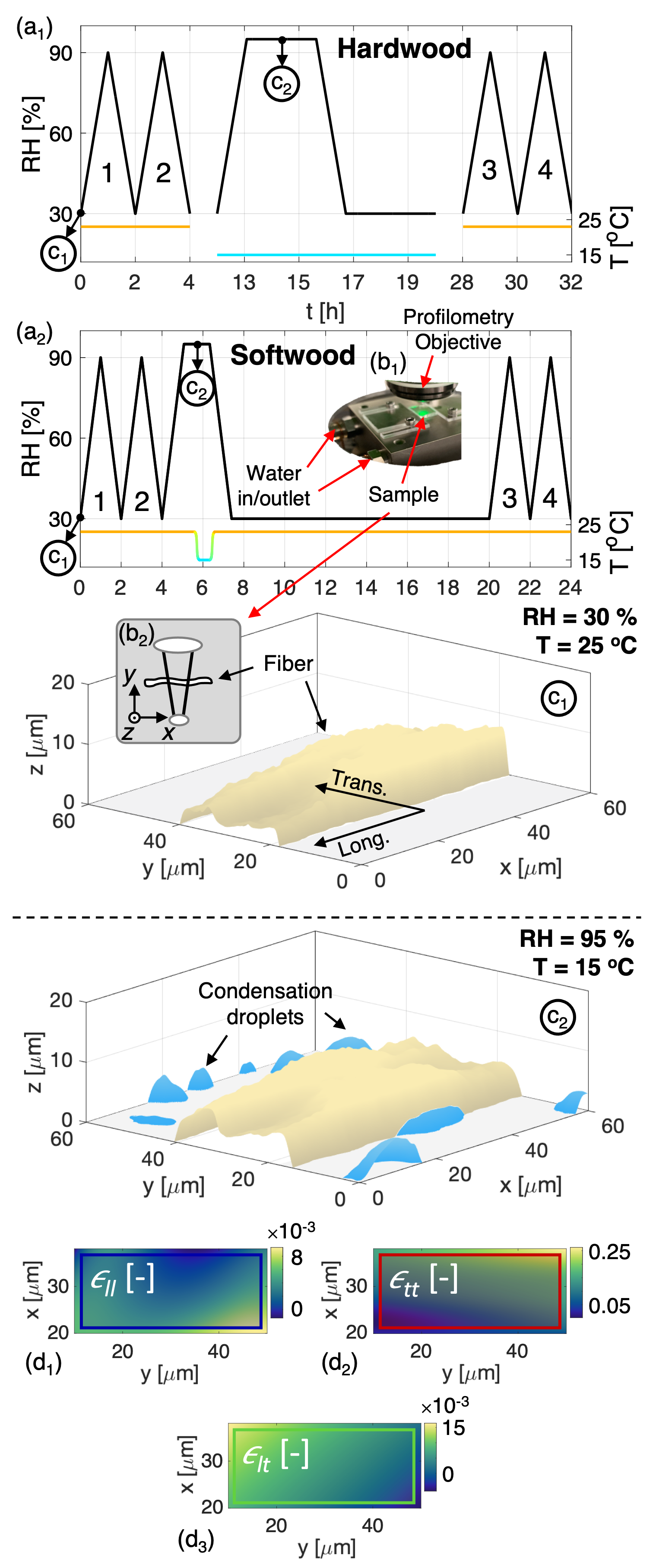}
	\caption{Single fiber hygro-expansion test (extended from \citep{vonk2020robust}: a fiber is (a) exposed to a combined RH an $T$ trajectory (which is slightly different for (a\textsubscript{1}) HW and (a\textsubscript{2}) SW fibers), and (b) the fiber is clamped with two nylon threads (b\textsubscript{1}) for maximum freedom while the region-of-interest remains in the field-of-view of the optical profilometer (b\textsubscript{2}), which (c) is used to monitor the surface topography starting from the (c\textsubscript{1}) dry state to (c\textsubscript{2}) the fully wet state (with coloring of fibers and droplets based on the topography) and vice versa, and (d) the topographies are correlated using GDHC to yield the evolution of the (d\textsubscript{1}) longitudinal ($\epsilon_{ll}$), (d\textsubscript{2}) transverse ($\epsilon_{tt}$), and (d\textsubscript{3}) shear strain field ($\epsilon_{lt}$) shown here for the strain accumulated during the second wetting slope from 30 to 90\% RH.}
	\label{fig:fiber_method}
\end{figure}

From the handsheets a total of, respectively, five FD and twelve RD HW fibers and ten FD and ten RD SW fibers were extracted by means of delaminating the paper sheet and cutting the naturally sticking out fibers \citep{hirn2006investigating}. For these sample fibers, only fibers without any noticeable pre-deformation or damage were accepted. Moreover a comprehensive analysis on 100 fibers picked from the handsheets showed that the morphology of the fibers was completely random in terms of straightness, twist, etc. confirming that the fiber extraction procedure did not select specific fibers, and consequently, the fibers tested are also of random nature. For the tests, each fiber is delicately clamped using two nylon threads following the method proposed in \citep{vonk2020robust}, as schematically shown in Figure \ref{fig:fiber_method} (b2). To enable GDHC \citep{vonk2020robust}, a micro-particle pattern was applied by exposing the fiber to a mist of freely floating sub-micro particles, created with a mystification setup \citep{shafqat2021cool}. 
\noindent

\subsection{Single fiber hygro-expansivity experiments}
The fibers (one by one) were subsequently tested inside a climate chamber underneath an optical profilometer, in which the RH is varied using a RH sensor (feedback control) located approximately 4 cm from the fiber (closer is not possible to avoid collision), during which the 3D deformation of the fiber is captured (one topography per 30 s). The fiber is clamped on a flow-through element (Figure \ref{fig:fiber_method} (b\textsubscript{2})) which is connected to an external water temperature regulator enabling temperature control of the specimen. Examples of \textit{in-situ} acquired topographies in the dry and wet state are shown in Figure \ref{fig:fiber_method} (c). After testing, GDHC, a topography correlation algorithm \citep{vonk2020robust}, is used to obtain the full-field fiber deformation, i.e. the longitudinal ($\epsilon_{ll}$), transverse ($\epsilon_{tt}$) and shear hygro-expansion ($\epsilon_{lt}$) field shown in Figure \ref{fig:fiber_method} (d) for the deformation accumulated during the second wetting slope (between 30 and 90\% RH) of a SW fiber. The strain fields are subsequently averaged towards a single scalar value of $\epsilon_{ll}$, $\epsilon_{tt}$, and $\epsilon_{lt}$, for each topography in time, enabling visualization of the evolution of the hygro-expansion. \\ \indent
Figure \ref{fig:fiber_method} shows that two slightly different RH/$T$ trajectories are used to test fibers from each wood type. In both trajectories the fiber is first subjected to (i) two linearly increasing RH cycles from 30$-$90$-$30\% (cycle 1$-$2) and temperature of 23\textsuperscript{o}C, then (ii) a wetting cycle during which the fiber's moisture content is maximized by means of lowering the specimen temperature to 15\textsuperscript{o}C, (locally) increasing the RH \citep{fellers2007interaction}, and generating condensation droplets as visible in Figure \ref{fig:fiber_method} (c\textsubscript{2}), and finally (iii) two 30$-$90$-$30\% RH cycles (cycle 3$-$4) at a temperature of 23\textsuperscript{o}C. Each change in RH set-point is conducted with a slope 1\%/min, equal to that of the handsheets. This strategy is chosen to test if an initially RD fiber is able to "transform" into a FD fiber when subjected to sufficient moisture.\\ \indent
The wetting cycle was, however, performed differently for HW and SW. The HW fibers were tested first, and the initial wetting cycle strategy was optimized. However, as will be elaborated below, the SW fibers exhibited significantly different hygro-mechanical behavior during the wetting cycle than the HW fibers. In order to understand these differences, higher moisture content levels were required, hence the approach was adapted. Additionally, the adapted approach enabled shorter total test times. For HW, the temperature is altered before and after the wetting cycle and kept constant for 8 hours to reach full equilibrium, which is characterized by the steps in Figure \ref{fig:fiber_method} (a\textsubscript{1}). For SW, the temperature is lowered when the RH has been 95\% for 30 minutes, resulting in significantly larger amounts of condensation in a shorter time compared to the HW approach. The temperature is increased again, combined with linearly lowering the RH as annotated in Figure \ref{fig:fiber_method} (a\textsubscript{2}), when the fiber starts to be covered by water, making the GDHC impossible. Hence, in contrast to the HW approach, the temperature and RH trajectories of SW are not fixed and depend on the condensation formation. 

\subsection{Sheet-scale hygro-expansivity experiments}
\begin{figure}[]
	\centering
	\includegraphics[width=0.5\textwidth]{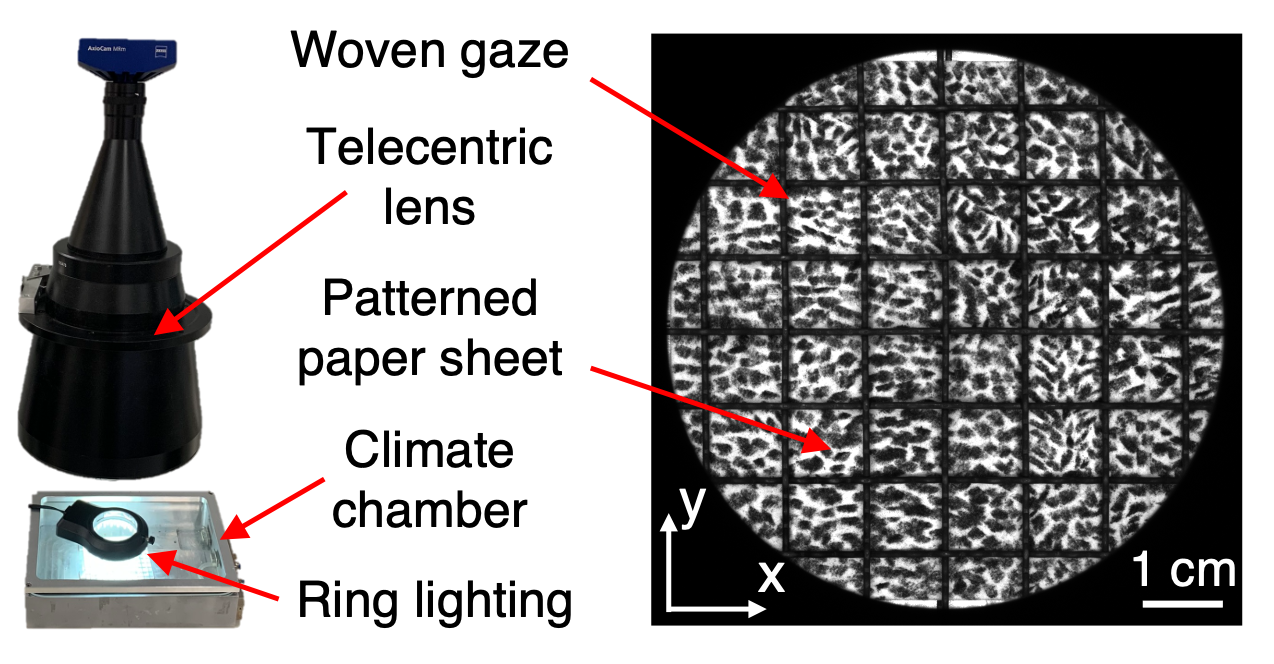}
	\caption{Paper sheet-scale hygro-expansion setup, consisting of a climate chamber underneath a telecentric lens-camera setup. This specific lens annihilates artificial strain due possible present out-of-plane deformations. The patterned paper sheet is placed in between two flat (\textsuperscript{$\sim$}5 $\mu$m corner-to-corner height deviation) woven gazes to minimize out-of-plane deformation to further minimize artificial strains.}
	\label{fig:sheet_method}
\end{figure}

The full-field in-plane paper hygro-expansion method based on GDIC involves placing the patterned paper sheet in between two flat (\textsuperscript{$\sim$}5 $\mu$m corner-to-corner height deviation) woven steel gazes, which are spaced with the paper thickness to minimize out-of-plane deformation, which is known to occur during sheet-scale hygro-expansion measurements \citep{kulachenko2005tension, de2017moisture}, shown in Figure \ref{fig:sheet_method}. This paper sheet is tested inside a climate chamber, in which the relative humidity (RH) is regulated, and captured using a telecentric lens-camera setup, as shown in Figure \ref{fig:sheet_method}. This setup enables minimization of artificial strains due to out-of-plane deformations, because of the lens' invariant magnification around the focus point. The RH is regulated by an external humidifier (\textit{Cellkraft P-10 series}). In the experiments, two RD and two FD handsheets were tested for HW and SW, with each test consisting of six linearly increasing RH cycles from 30$-$90$-$30\%, with a slope of 1\%/min, resulting in a total duration of 12 hours. All (eight) sheets were acclimatized at 30\% RH for at least 12 hours before testing. Images were captured once per two minutes and were correlated using a GDIC framework with linear shape functions to find the linearly varying evolution of the displacement field corresponding to constant strain fields $\epsilon_{xx}(x,y) = \epsilon_{xx}$, $\epsilon_{yy}(x,y) = \epsilon_{yy}$, and $\epsilon_{xy}(x,y) = \epsilon_{xy}$ \citep{neggers2016image}. As $\epsilon_{xy}$ remains zero and $\epsilon_{xx}$ equals $\epsilon_{yy}$ within uncertainty margins, the sheet hygro-expansion is computed as $\epsilon_s = (\epsilon_{xx}+\epsilon_{yy})/2$. Note that GDIC is a 2D formulation, while Global Digital Height Correlation (GDHC), which is used for the fiber hygro-expansivity below, is a quasi-3D framework. Finally, local DIC, which is more common (also commercially available), has been used earlier for multiple paper mechanics problems, e.g., paper sheet-scale hygro-expansion measurements \citep{fellers2007interaction} and strain field evolution of paper sheets subjected to bi-axial tension \citep{alzweighi2022evaluation}. Here GDIC is adopted as it is more accurate for low spatial variations in the strain field. 

\section{Results and discussion}
The produced RD handsheets exhibited a significantly lower hygro-expansion than the FD handsheets, as will be shown below in Figure \ref{fig:sheet_test}, similar to other work in the literature, indicating that the handsheet fabrication process was conducted as expected. The fiber hygro-expansion is elaborated first because this reveals the most novel and interesting results in terms of hygro-expansivity, which will be later on compared to the sheet-scale hygro-expansion.

\subsection{Evolution of the fiber hygro-expansivity}
\begin{figure}[]
	\centering
	\includegraphics[width=\textwidth]{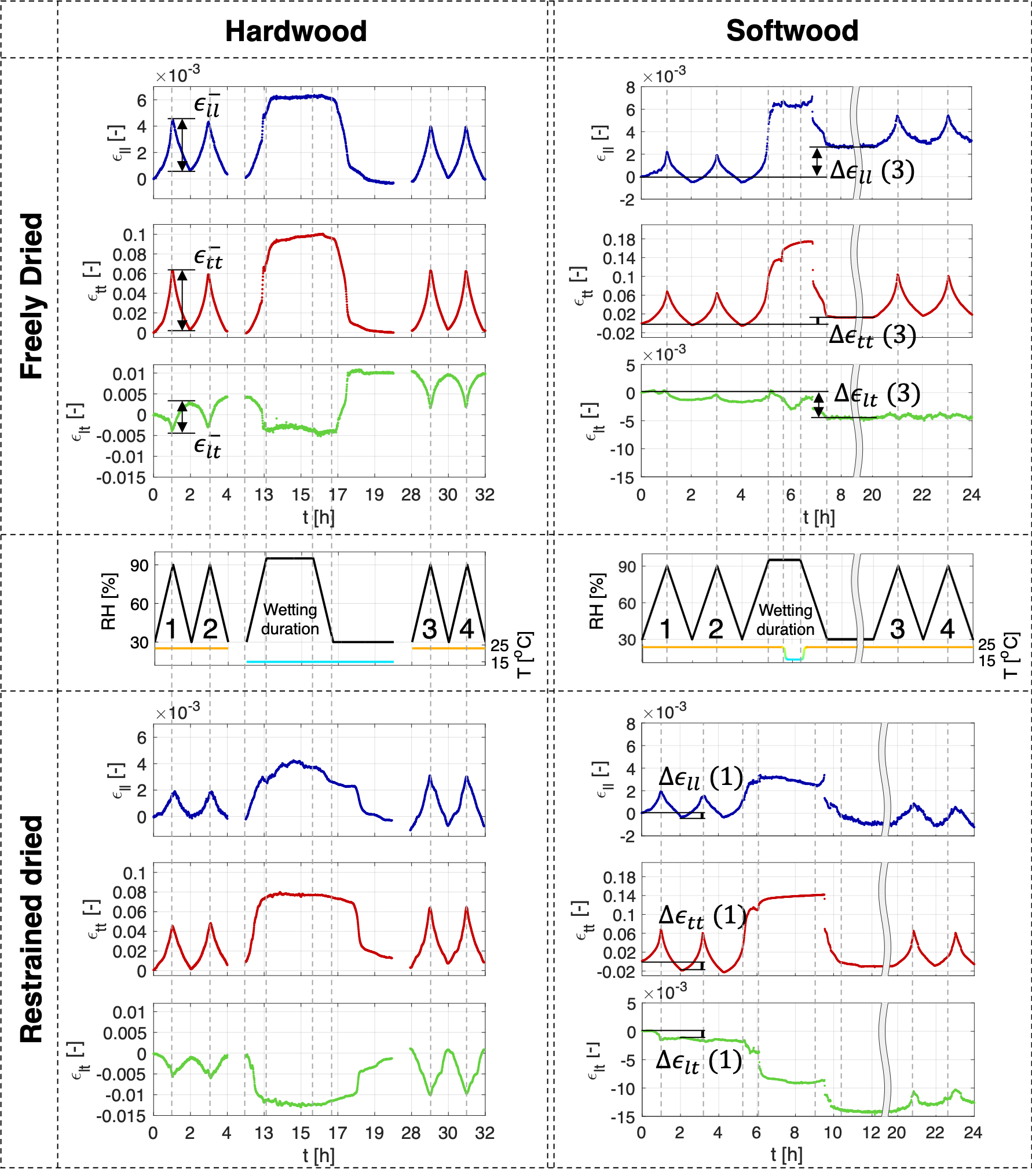}
	\caption{The longitudinal ($\epsilon_{ll}$), transverse ($\epsilon_{tt}$) and shear hygro-expansion ($\epsilon_{lt}$) evolution of a typical SW and HW, RD and FD fiber. The delay between the hygro-expansion and the RH is due to the presence of condensation. The cooling time during the wetting cycle of the RD SW fiber was slightly longer than the FD SW fiber because the condensation formation took longer. The release of dried-in strain ($\Delta\epsilon$) between the start and (i) after the first RH cycle (1, in the RD SW curve), and (ii) just before the third RH cycle, after the wetting cycle (3, in the FD SW curve) are extracted and shown in Figure \ref{fig:DI}. The change in strain for each drying slope from 90 to 30\% RH ($\epsilon^-_{ll}$, $\epsilon^-_{tt}$, and $\epsilon^-_{lt}$ depicted in the FD HW curve) is extracted for each RH cycle for further analysis.}
	\label{fig:trend_fiber}
\end{figure}

The hygro-expansion response of a typical FD and RD, HW and SW fiber is given in Figure \ref{fig:trend_fiber}. The strong anisotropic swelling behavior of the fibers is directly visible. Optimization of the fiber preparation and testing method have resulted in a significantly better strain resolution compared to \cite{vonk2020robust}, whereby the precision improved from 1$\cdot$10\textsuperscript{-4} and 7$\cdot$10\textsuperscript{-4} in longitudinal and transverse direction respectively, to 1$\cdot$10\textsuperscript{-4} and 2$\cdot$10\textsuperscript{-4} \citep{vonk2023frc}. Furthermore, some fibers (e.g., FD SW fiber) show scattered data during the wetting cycle, which is caused by severe condensation occurring at the fiber surface (and possibly the particles), making the GDHC unstable. Nevertheless, the global curve is still visible and reliable, because all final solutions converged properly. Note that the fiber method enables monitoring the surface strain field during the transition to an almost fully wet surface, during the wetting cycle itself, and during the reversed transition to a dry surface (which will be studied in more detail in \citep{vonk2023bonds}. \\ \indent 
Interestingly, at the end of the wetting cycle, when the RH is decreasing, the hygro-expansion is lagging behind the RH decrease (most prominently visible for the HW fibers) due to the severe condensation present around the fiber, requiring time to evaporate. The condensation around the fiber results in an unknown RH value near the fiber, which is not a concern because the purpose of this wetting cycle is to characterize the release of dried-in strain before and after this period, for which a long period of high moisture content levels is required. The moment at which the condensation initiates shows a rather high variability for the original RH/$T$ procedure applied to the HW fibers. For the adapted RH/$T$ procedure applied to the SW fibers, no droplet formation is visible during the 30 minutes at 95\% RH and high temperature, but lowering the stage temperature immediately initiates droplet formation and consequently increases the fibers' moisture content, characterized by the rise in $\epsilon_{tt}$ (also slight increase in $\epsilon_{ll}$) of both SW fibers in Figure \ref{fig:trend_fiber}. This is the moment where the fiber exhibits hydro-expansion instead of the initially studied hygro-expansion \citep{larsson2009influence}. The strength of the approach lies in the fact that the hygro- and hydro-expansion of the same type of fiber within the same experiment can be studied. For almost all fibers, the hydro-expansivity entails significantly larger strains compared to the hygro-expansion induced during RH cycles 1$-$2, except for a few fibers exhibiting severe shrinkage during the wetting period, confirming that a higher moisture content is reached. All fibers tend to reach an equilibrium during the wetting cycle (although some small, random variations in the order of \textsuperscript{$\sim$}10\textsuperscript{-3} strain remain, showing that the strain accuracy is approximately one order of magnitude lower for a full wet surface compared to a dry surface at lower RH levels), implying saturation of the moisture content and full release of dried-in strain. The transient fiber hydro-expansion during the wetting cycle is studied in more detail in \citep{vonk2023hydro}.

\subsubsection{Release of dried-in strain}
\begin{figure}[t!]
	\centering
	\includegraphics[width=\textwidth,trim=6 6 6 4,clip]{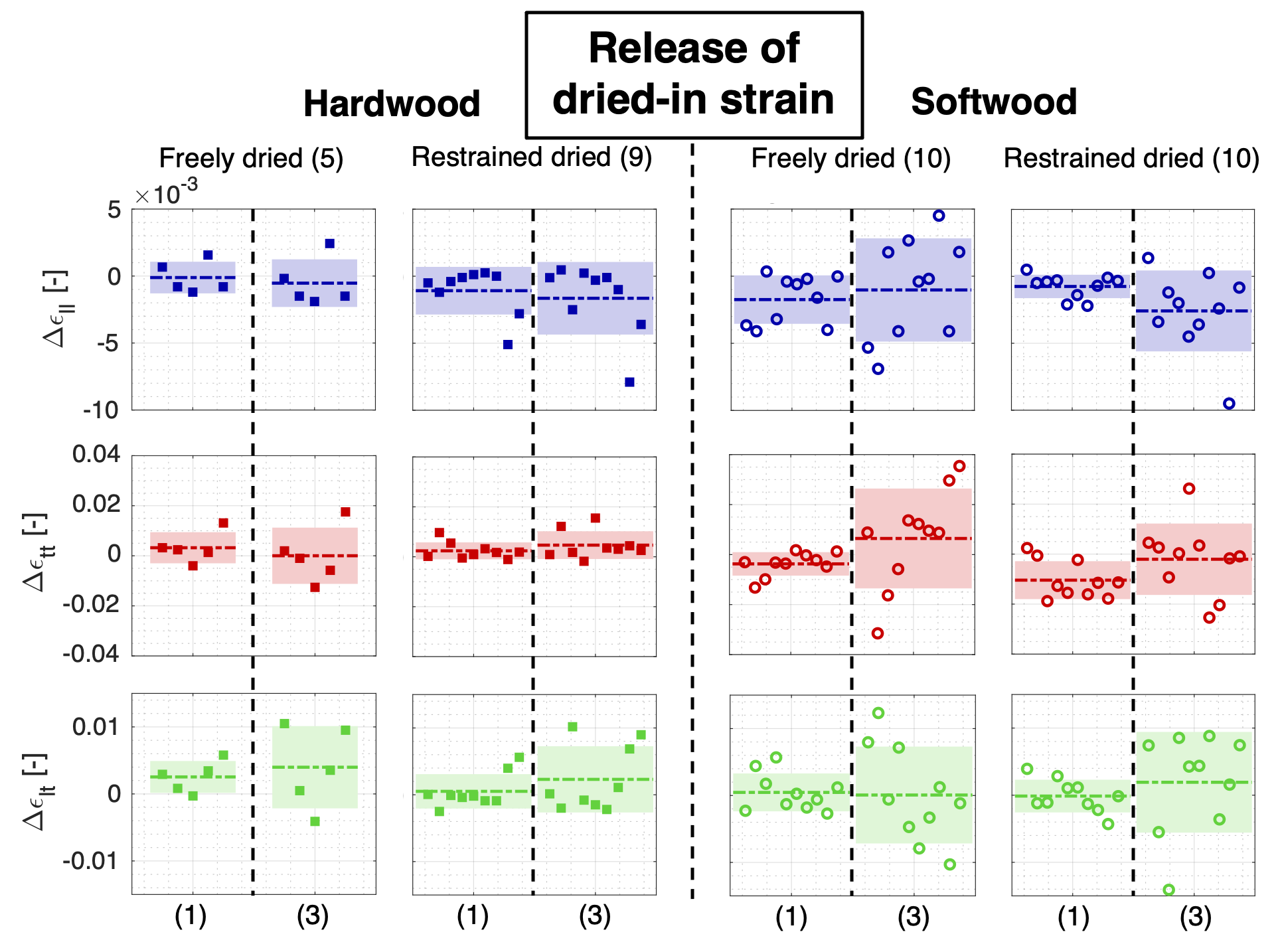}
	\caption{The release of dried-in strain in longitudinal ($\epsilon_{ll}$), transverse ($\epsilon_{tt}$), and shear direction ($\epsilon_{lt}$) after the first RH cycle, labeled (1) in Figure \ref{fig:trend_fiber}, and just before the third cycle, after the wetting cycle, labeled (3) in Figure \ref{fig:trend_fiber}. The values are separately given per fiber, together with the average and standard deviation of all fibers, represented by the dashed line and colored band respectively.}
	\label{fig:DI}
\end{figure}

All fibers show a clear release of dried-in strain during the wetting slope of RH cycle 1, because it deviates from the subsequent wetting slopes, similar to earlier single fiber hygro-expansion measurements \citep{vonk2021full, vonk2020robust, vonk2023frc} and earlier sheet-scale hygro-expansion works \citep{larsson2008influence, urstoger2020microstructure, uesaka1992characterization, vonk2023frc}, as well as the sheet-scale experiments shown later in Figure \ref{fig:sheet_test}. More interestingly, all fibers show a release of dried-in strain after the wetting cycle, visible in the fact that for each fiber at least one of the three strain components shows a large difference between the start of the wetting cycle and the start of RH cycle 3 (at which the temperature and RH are the same as the start of the experiment). The average release of dried-in strain in $\epsilon_{ll}$, $\epsilon_{tt}$, and $\epsilon_{lt}$, i.e., $\Delta\epsilon_{ll}$, $\Delta\epsilon_{tt}$, and $\Delta\epsilon_{lt}$ after RH cycle 1, labeled (1), and between the start and just before RH cycle 3, labeled (3), considering all fibers, are presented in Figure \ref{fig:DI}. \\ \indent
The values per fiber are given separately, and the average and standard deviation of all fibers visualize the global trend. Clearly, not all dried-in strains are released after the first RH cycle (1), because the dried-in strain after the wetting cycle (3) is on average larger. Hence, dried-in strains are still stored inside the fiber after being subjected to 90\% RH. Furthermore, all groups of fibers show both positive and negative $\Delta\epsilon_{ll}$ values, implying that the fibers inside both FD and RD handsheets experienced drying under compression or tension. However, more fibers inside the FD handsheets were dried under compression characterized by the larger number of positive dried-in strain values. These results are in line with \cite{nanko1995mechanisms}, who showed that non-bonded fiber segments at the paper surface experienced either swelling or shrinkage during FD and RD depending on their neighboring fibers, and the FD case showed more drying under compression. Furthermore, the larger $\Delta\epsilon_{ll}$ for RD compared to FD is attributed to the sheet dryer exerting an out-of-plane compression on the sheet. Thereby imposing a drying (tensile) stress on most fibers, resulting in a larger portion of fibers being dried under tensile stresses. In contrast, during FD, so-called "micro-compressions" may induce in the bonded areas \citep{page1962new}. The scatter in $\Delta\epsilon_{ll}$ is also visible in $\Delta\epsilon_{tt}$ (mainly for SW), suggesting that the applied tensile or compressive stress during drying, respectively, contracts or expands the cross-section of the fibers. \\ \indent
To investigate how the release of dried-in strains and the drying procedure affects the hygro-expansivity of the fibers, the shrinkage before the wetting cycle (i.e. RH cycles 1$-$2) and after the wetting cycle (i.e. RH cycles 3$-$4) are extracted and discussed in the following.

\subsection{Comparison of hygro-expansion magnitudes}
\begin{figure}[t!]
	\centering
	\includegraphics[width=0.5\textwidth]{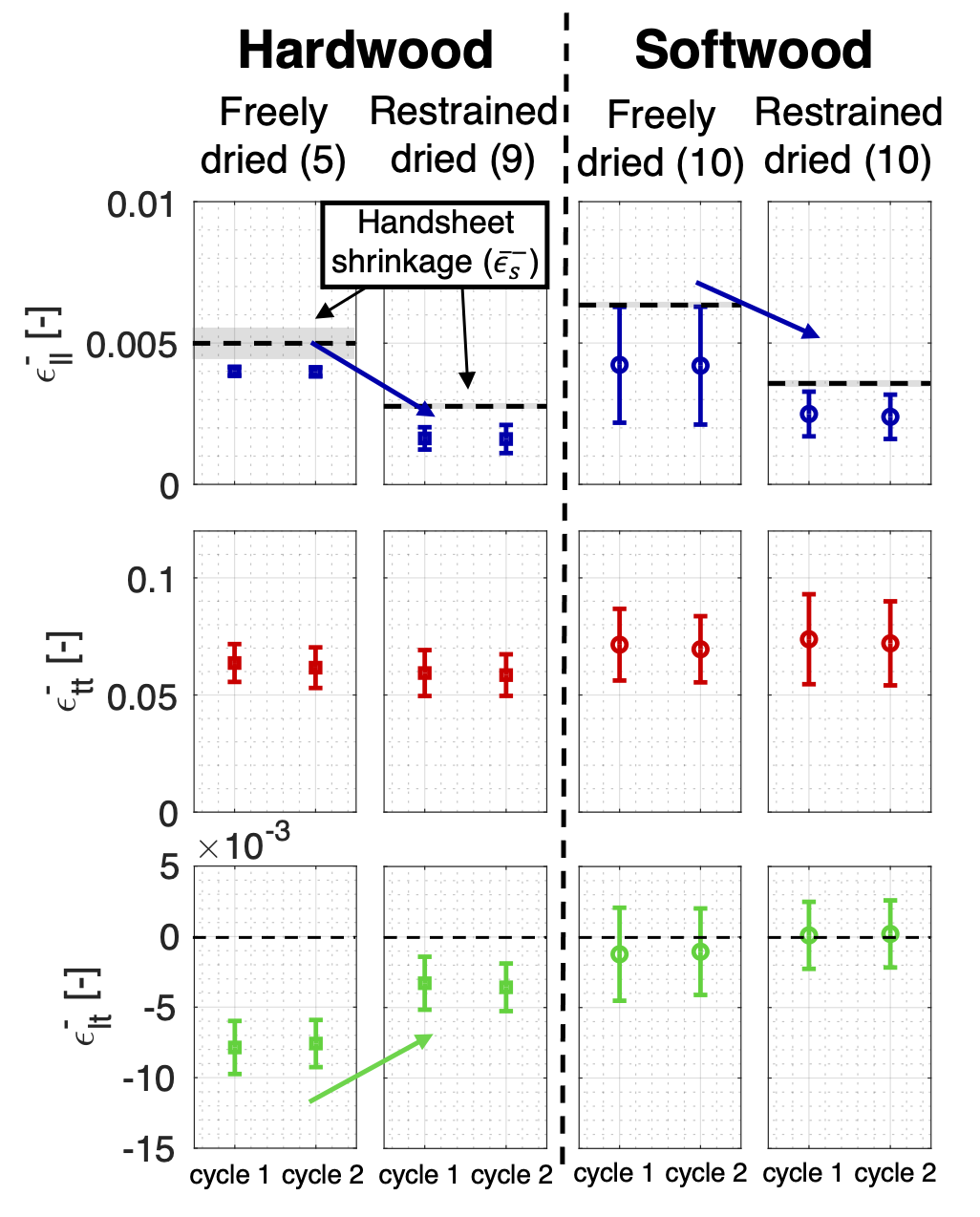}
	\caption{The average longitudinal ($\epsilon^-_{ll}$) and transverse shrinkage ($\epsilon^-_{tt}$), and shear strain change ($\epsilon^-_{lt}$) during the drying slopes from 90 to 30\% RH for cycles 1$-$2, as annotated for RH cycle 1 of the FD HW fiber in Figure \ref{fig:trend_fiber}, of all HW (five and nine fibers from the FD and RD HW handsheets respectively) and SW fibers (ten fibers each from the FD and RD SW handsheets), i.e. $\bar{\epsilon}^-_{ll}$, $\bar{\epsilon}^-_{tt}$, and $\bar{\epsilon}^-_{lt}$, including standard deviation. The total average handsheet shrinkage ($\bar{\epsilon}^-_s$, given in Table \ref{tab:sheet_strain}) is added to the $\bar{\epsilon}^-_{ll}$ plot.}	
	\label{fig:strains_1_2}
\end{figure}

The average (with standard deviation) longitudinal and transverse shrinkage, and shear strain change of all FD and RD HW and SW fibers during the drying slope from 90 to 30\% RH for cycles 1$-$2 (before the wetting cycle), i.e. $\bar{\epsilon}^-_{ll}$, $\bar{\epsilon}^-_{tt}$, and $\bar{\epsilon}^-_{lt}$ are shown in Figure \ref{fig:strains_1_2}. For better analysis, the shrinkage is defined as a positive strain value, i.e. the strain changes are obtained by subtracting the value at 90\% by the value at 30\%, as annotated for RH cycle 1 of FD HW in Figure \ref{fig:trend_fiber}. Which results, for this specific hygro-expansion curve, in positive values (shrinkage) for $\epsilon^-_{ll}$ and $\epsilon^-_{tt}$, and negative $\epsilon^-_{lt}$. Subsection \ref{sec:app1} provides (i) the strains per fiber for each cycle separately given in Figures \ref{fig:app_hw} (a) and \ref{fig:app_sw} (a), of which marker numbering is consistent with the dried-in strain releases given in Figure \ref{fig:DI}, (ii) the total average shrinkage considering cycles 1$-$2 of all fibers, and (iii) an explanation why three RD HW fibers are excluded from the analysis and extra details on RD HW fibers which exhibited severe shrinkage during wetting. \\ \indent
All fibers show a low cycle-to-cycle variability when comparing RH cycle 1 to 2, whereas only the HW fibers demonstrate a low fiber-to-fiber variability, similar to results in \citep{vonk2021full}. The latter is attributed to the significantly larger MFA range of the SW compared to the HW fibers \citep{french2000effect, barnett2004cellulose, cown2004wood, donaldson2008microfibril}, which is known to correlate to the fiber's hygro-expansivity \citep{meylan1972influence, yamamoto2001model}. Interestingly, for the SW fibers, the scatter of $\bar{\epsilon}^-_{ll}$ is lower for the RD SW fibers, opposite to what can be expected, which is discussed later. \\ \indent 
Regarding the shrinkage, both for HW and SW, the FD fibers show a larger shrinkage than the RD fibers, especially in longitudinal direction. For HW and SW, the RD procedure mainly affects the longitudinal shrinkage and, to a lesser extent, also the shear, while not affecting the transverse shrinkage. On average, the FD HW fibers shrink 2.5 times more in longitudinal and 1.1$\pm$0.3 times more in transverse direction compared to RD HW fibers, and the shear is 2.3 times larger. The FD SW fibers shrink 1.8 times more in longitudinal and similarly (1.0) in transverse direction than the RD SW fibers, and the shear is 5.5 times less. The large scatter of $\bar{\epsilon}^-_{lt}$ of the SW fibers stems from the RD shear strains being close to zero. Furthermore, the SW fibers exhibit a slightly larger longitudinal and transverse shrinkage than the HW fibers, which is in line with the slightly larger sheet-scale shrinkage of SW compared to HW described in Subsection \ref{sec:sheet} below, and the literature \citep{nanko1995mechanisms, uesaka1997effects}. \\ \indent
Collectively, because the hygro-expansivities (shrinkage) of the RD and FD fibers are significantly different, it can be expected that the fiber structure itself is different between the RD and FD fibers, which directly affects the sheet hygro-expansion. This is visualized by comparing $\bar{\epsilon}^-_{ll}$ to the average handsheet shrinkage, as is done in the $\bar{\epsilon}^-_{ll}$ plot in Figure \ref{fig:strains_1_2}. Hence, to explain the sheet hygro-expansion differences, it is not needed to introduce geometrical differences that have been hypothesized to occur in the bonded regions \citep{uesaka1994general}, as earlier questioned by \cite{urstoger2020microstructure}, supporting the earlier works stating that structural fiber differences are sufficient to explain the sheet-scale differences \citep{nanko1995mechanisms, jentzen1964effect, van1961some}. The structural differences and the driving mechanisms of these hygro-expansivity differences are studied next, where first only HW is considered.

\subsubsection{Hardwood fibers: driving mechanism of the hygro-expansivity difference} 
\begin{figure}[t!]
	\centering
	\includegraphics[width=\textwidth,trim=6 4 4 4,clip]{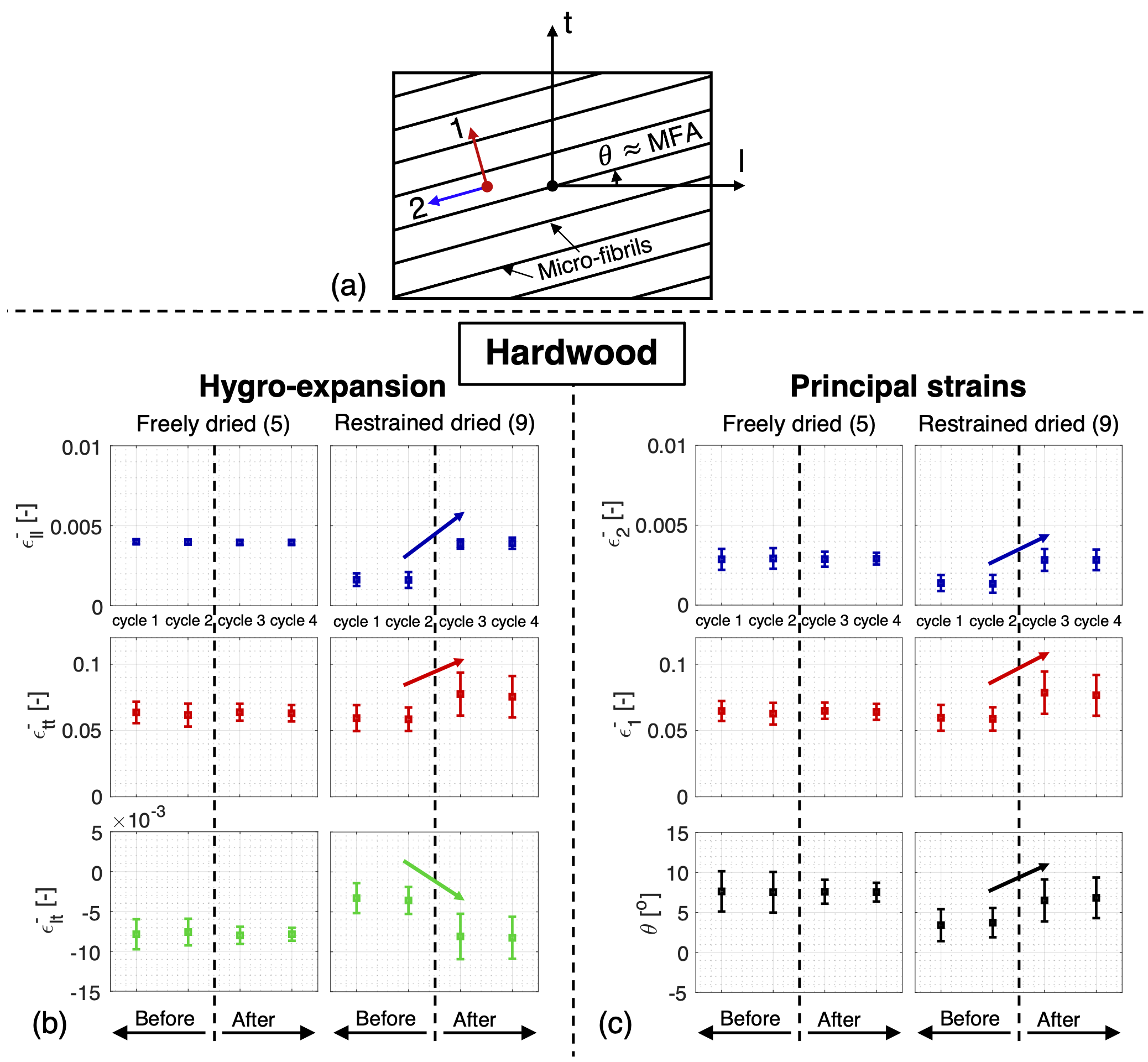}
	\caption{(a) Schematic visualization of the principal strains directions with 1 oriented in the direction of the major strain, i.e., perpendicular to the fibril orientation and 2 for the minor strain oriented along the micro-fibrils, and the major-minor strain angle ($\theta$) related to the MFA. (b) The average longitudinal and transverse shrinkage, and shear strain change during the drying slopes from 90 to 30\% RH per RH cycle of all HW fibers (five and nine fibers extracted from the FD and RD HW handsheets respectively), i.e. $\bar{\epsilon}^-_{ll}$, $\bar{\epsilon}^-_{tt}$, and $\bar{\epsilon}^-_{lt}$, including standard deviation. (c) The average principal strains and major minor strain angle, i.e. $\bar{\epsilon}^-_1$ and $\bar{\epsilon}^-_2$, and $\bar{\theta}$ are computed from the full-field hygro-expansion, of which $\theta$ approximates the MFA.}
	\label{fig:strains_hw}
\end{figure}

\cite{jentzen1964effect} showed that the MFA of RD fibers is lower than FD fibers. Therefore, it would be insightful if the MFA of each of the above-tested FD and RD, HW and SW fibers can be measured or assessed. Unfortunately, most MFA measurement techniques in literature (i) have insufficient accuracy to detect the small differences in MFA between the FD and RD fibers which are expected to be low, as will be shown later on, while the MFA is known to vary along the fiber length \citep{anagnost2002variation, ye1994microscopic}, and/or (ii) require wood slices \citep{donaldson2008microfibril, cave1997theory, wang2001improved, saren2006determination}. Only one method (proposed by \cite{heinemann2014microfibril}) based on transmission white-light polarization microscopy has been proposed to measure the MFA of individual fibers, which is explored. The method was adopted and further optimized, and now allows determination of the MFA of the full fiber surface instead of a few discrete locations, of which a detailed description and an example figure is given in Subsection \ref{app:mfa}. Unfortunately it was concluded that the method is fundamentally not applicable to paper fibers, because the polarized light travels though the front and back fiber wall of which the micro-fibrils are crossed due to their helical structure, inevitably resulting in an MFA around 0\textsuperscript{o}, as explained in more detail in Subsection \ref{app:mfa}. Nevertheless, as will be shown next the MFA can indirectly be deduced from the full-field fiber hygro-expansion measurements.\\ \indent
In \citep{vonk2021full} it was demonstrated that computation of the principal strains, by means of an eigenvalue decomposition from the full-field strain tensor with components $\epsilon^-_{ll}$, $\epsilon^-_{tt}$, and $\epsilon^-_{lt}$, enables the determination of not only the major and minor strain ($\epsilon_{1}$, and $\epsilon_{2}$), but also the in-plane principal strain direction, i.e. the major-minor strain angle ($\theta$). Considering (i) that the fiber's mechanical behavior is dominated by the S2 layer, and (ii) that the direction of major strain will be perpendicular to the relatively stiff micro-fibrils, due to the deformation induced by the swelling of the hemi-celluloses in between the micro-fibrils, the MFA has to be (close to) perpendicular to the major-minor strain angle, as shown in Figure \ref{fig:strains_hw} (a). Hence, the average major and minor strain, and the average major-minor strain angle during shrinkage from 90 to 30\% RH, i.e. respectively $\bar{\epsilon}^-_1$ and $\bar{\epsilon}^-_2$, and $\bar{\theta}$ are given for all RH cycles along with the regular average strains ($\bar{\epsilon}^-_{ll}$, $\bar{\epsilon}^-_{tt}$, and $\bar{\epsilon}^-_{lt}$) in Figure \ref{fig:strains_hw}. The shrinkage and principal strains of all fibers separately are again given in Figure \ref{fig:app_hw}. \\ \indent
Let us first examine RH cycles 1$-$2, before the wetting cycle. Clearly, all FD and RD HW fibers are within the MFA range reported in the literature, i.e. 0$-$11\textsuperscript{o} \citep{french2000effect, donaldson2008microfibril}. Furthermore, the HW fibers show distinctly different values for $\theta$, i.e., 7.57$\pm$2.67 and 3.56$\pm$2.92\textsuperscript{o} between FD to RD respectively. This shows that the MFA of the RD HW fibers is indeed lower than the FD fibers, which is in agreement with the findings of \cite{jentzen1964effect}.\\ \indent
The differences may be rationalized by analyzing the principal strains in Figure \ref{fig:strains_hw} (c). $\bar{\epsilon}^-_2$ of the FD and RD HW fibers shows that the shrinkage along the fibril length is actually non-zero, indicating that the micro-fibrils themselves can swell or shrink in their longitudinal direction, suggesting that the "dislocated regions" in the cellulose micro-fibrils are indeed accessible to water \citep{agarwal2016probing}. $\bar{\epsilon}^-_2$ of the RD HW fibers (RH cycles 1$-$2) is significantly lower than the FD HW fibers, indicating that the "dislocated cellulose regions" shrink less, mainly affecting the fiber's longitudinal shrinkage due to HW's low MFA \citep{french2000effect, donaldson2008microfibril}. These results are in good agreement with the theory of \cite{salmen1987development} who stated, for RD paper, that the "dislocated regions" in the cellulose micro-fibrils are stretched when dried under tension (inducing a residual tensile stress in the direction of the micro-fibrils). This in turn minimizes the swelling of these regions upon wetting, and therefore they contribute less to the fibers' overall hygro-expansivity. For FD fibers, the "dislocated regions" are dried without tension, and thus shrink upon drying, through which they can contribute more to the fiber's hygro-expansivity. Additionally, \cite{kulachenko2012elastic} showed that the drying procedure also affected the hygro-expansivity of nano-cellulose paper. Which consists of squared cross-sectional nano-fibrils with an height and width of 20 nm, corresponding to the size of nano-fibril aggregates that form the cell wall of paper fibers \citep{fahlen2005pore}. The reported difference by \cite{kulachenko2012elastic} between the FD and RD nano-paper's drying slopes from 80 to 20\% RH was \textsuperscript{$\sim$}0.20\% strain (RD: 0.75\% FD: 0.95\%), which is similar to the difference in $\bar{\epsilon}_{2}$ found in Figure \ref{fig:strains_hw} (c) between FD and RD (for an RH change of 90 to 30\%). This suggests that this hygro-expansion difference of the nano-cellulosic paper is driven by the swelling capabilities of the "dislocated regions" in the cellulose nano-fibrils as proposed by \cite{salmen1987development}. \\ \indent
The fact that the RD and FD HW fibers show distinctly different values of $\bar{\theta}$, indicates that the MFA of FD fibers is higher than RD fibers, hence constitutes a driving mechanism behind the difference in hygro-expansivity of the HW fibers, similar to the findings of \cite{jentzen1964effect}. Furthermore, the significant difference in $\bar{\epsilon}_2$ between the RD and FD HW fibers is consistent with the theory by \cite{salmen1987development}.\\ \indent 
To investigate if the release of dried-in strain is reversible, the shrinkage after the wetting cycle, i.e. RH cycles 3$-$4 given in Figure \ref{fig:strains_hw} is studied. When comparing RH cycles 1$-$2 to cycles 3$-$4, all FD fibers tend to show a similar average values before and after the wetting cycle, although a slight decrease of the scatter of $\bar{\epsilon}^-_{lt}$ and $\theta$ is observed. The shrinkage of the RD HW fibers is similar to the FD HW fibers after the wetting cycle (only $\bar{\epsilon}^-_{tt}$ may be slightly larger for the RD HW fibers), indicating that RD HW fibers, when subjected to sufficient water, can "transform" into fibers that exhibit a hygro-expansivity similar to FD HW fibers, which is essential knowledge for paper recyclability. This is, however, in contrast to the experiments conducted in \citep{vonk2023frc}, during which the RD HW fibers were subjected for a long duration to an RH level of 95\% but without cooling, resulting in the fiber surface not becoming fully wet. Consequently, the fibers did not completely "transform" to FD fibers, showing the importance of cooling down the specimen to enforce higher moisture content levels of the fibers in order to activate the release of the fiber's dried-in strain. Furthermore, the RD HW fibers in Figure \ref{fig:strains_hw} (c) yield a $\theta$ of 6.66$\pm$2.34\textsuperscript{o} for RH cycles 3$-$4, which is equal to RH cycles 1$-$2 of the FD HW fibers within error margins, indicating that the change in $\theta$ between the two drying procedures is close to reversible. Additionally, $\bar{\epsilon}^-_2$ of the RD HW fibers (RH cycles 3$-$4) in Figure \ref{fig:strains_hw} (c) is similar to $\bar{\epsilon}^-_2$ of the FD HW fibers, indicating that the shrinkage along the fibril direction increased, giving further support to the theory of \cite{salmen1987development} which states that the difference in swelling capabilities of the "dislocated regions" in the cellulose micro-fibrils between FD and RD affects the fiber's hygro-expansivity. In short, the RD HW fibers can "transform" to fibers exhibiting the characteristics of FD HW fibers. \\ \indent
The SW fiber are studied next to investigate whether the above discussed theories and findings for HW fibers are also applicable to the SW fibers.

\subsection{Softwood fibers}
\begin{figure}[t!]
	\centering
	\includegraphics[width=\textwidth]{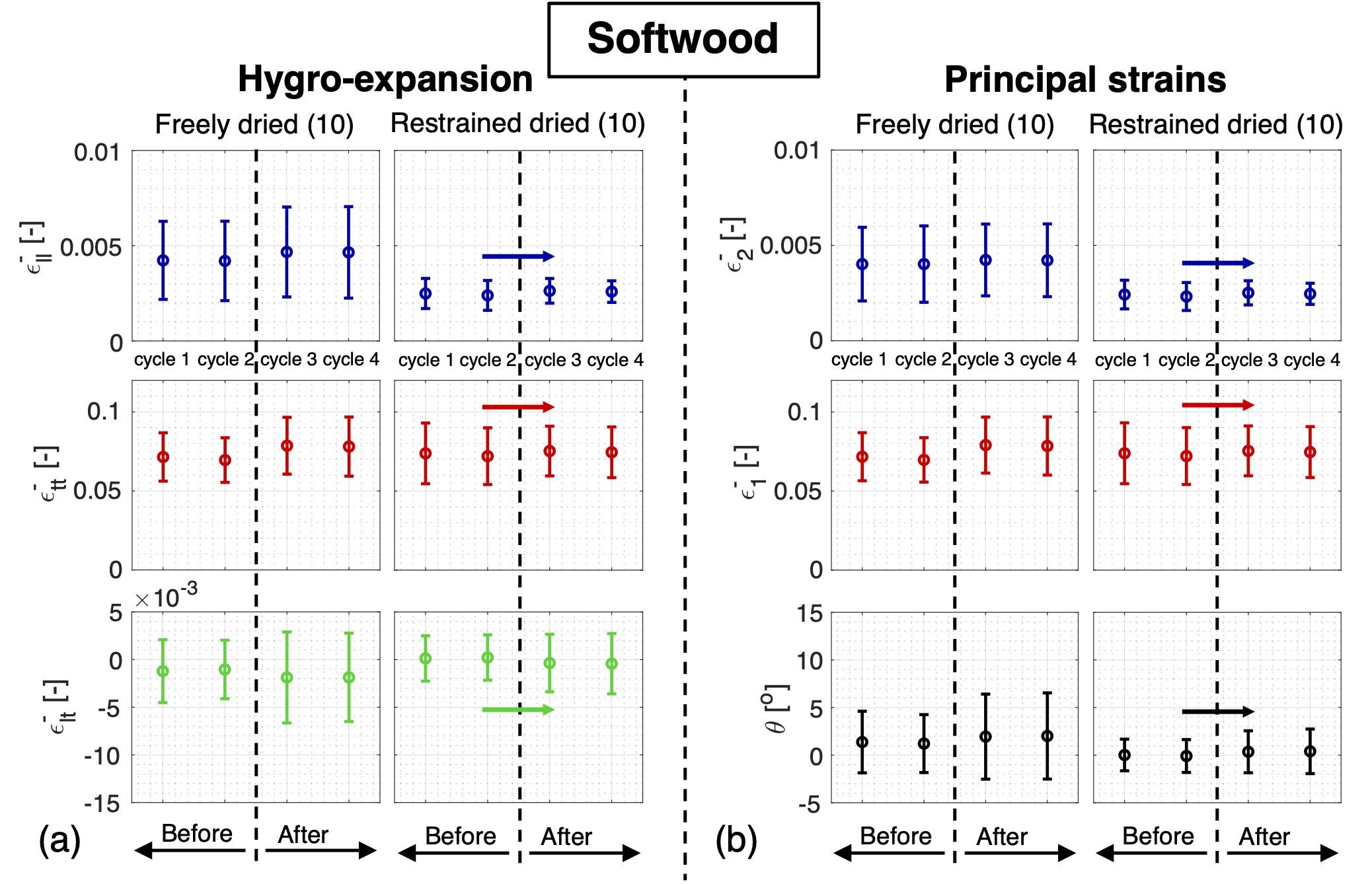}
	\caption{The average longitudinal and transverse shrinkage, and shear strain change during the drying slopes from 90 to 30\% RH per RH cycle of the SW fibers (ten fibers extracted from each FD and RD SW handsheet), i.e. $\bar{\epsilon}^-_{ll}$, $\bar{\epsilon}^-_{tt}$, and $\bar{\epsilon}^-_{lt}$, including standard deviation. The average principal strains ($\bar{\epsilon}^-_1$ and $\bar{\epsilon}^-_2$) and major-minor strain angle ($\bar{\theta}$) are computed from the full-field hygro-expansion, where the $\theta$ is (expected to approximate) the MFA.}
	\label{fig:strains_sw}
\end{figure}

The regular strains together with the principal strain computation of the SW fibers are given in Figure \ref{fig:strains_sw}. Let us first consider RH cycles 1$-$2. The SW fibers show both positive and negative values for $\theta$, which are outside of the MFA range reported in the literature, i.e. 8$-$39 \citep{barnett2004cellulose, cown2004wood}. This suggests that other mechanisms affect the direction of the deformation, which will be studied in the following. \\ \indent
Regarding the reversibility of the release of dried-in strain, remarkably, the shrinkage and principal strain components of the RD SW fibers in Figure \ref{fig:strains_sw} remain (nearly) unchanged after the wetting cycle (cycles 3$-$4). Thus, the RD SW fibers do not "transform" to fibers exhibiting the characteristics of FD SW fibers, in contrast to the HW fibers. The only difference in testing procedure between the HW and SW fibers was the duration of the wetting cycle, which was, most of the times, longer for the HW fibers, see Figure \ref{fig:fiber_method} (a). Therefore, the time dependence of the release of dried-in strain is investigated next.

\subsubsection{Moisture-induced "transformation" from restrained to freely dried fibers}
\begin{figure}[t!]
	\centering
	\includegraphics[width=0.5\textwidth]{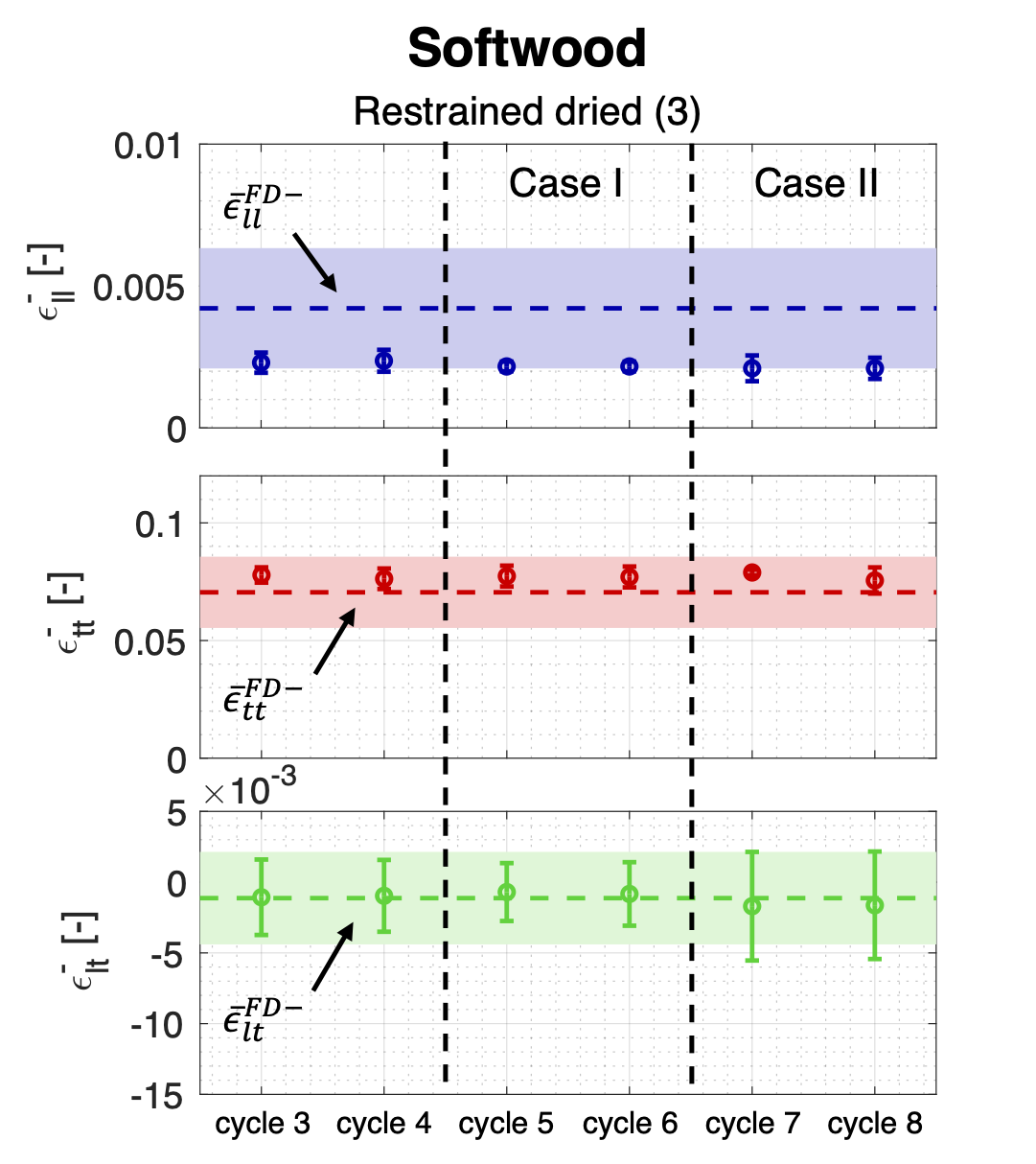}
	\caption{The change in $\bar{\epsilon}^-_{ll}$, $\bar{\epsilon}^-_{tt}$, and $\bar{\epsilon}^-_{lt}$ with standard deviation of three randomly selected RD SW fibers before and after soaking the fiber in water for 12 hours, while (i) being constrained by the nylon wires in water for 12 hours (Case I), and (ii) being completely free for 12 hours (Case II), also allowing the fiber to twist. $\bar{\epsilon}^-_{ll}$, $\bar{\epsilon}^-_{tt}$, and $\bar{\epsilon}^-_{lt}$ of the FD SW fibers considering RH cycles 1$-$2 including their standard deviation, as given in Table \ref{tab:fiber_strain}, are denoted by the horizontal dashed lines and bands.}
	\label{fig:change_hygro}
\end{figure}

A droplet of water (200 $\mu$L) was applied to three randomly selected RD SW fibers from the fibers shown in Figure \ref{fig:strains_sw}, and placed in a 100\% RH environment for 12 hours, to maintain the fiber fully soaked. Note that the fibers were still fixed on the glass substrate by the nylon wires. The fibers were afterwards again patterned (as most of the pattern is washed away when the water was removed) and subsequently subjected to two RH cycles of 30$-$90$-$30\% (cycles 5$-$6). The average strain change during drying from 90 to 30\%, i.e. $\bar{\epsilon}^-_{ll}$, $\bar{\epsilon}^-_{tt}$, and $\bar{\epsilon}^-_{lt}$ (with standard deviation) of RH cycles 3$-$4 (after the wetting cycle during the initial experiments) and RH cycles 5$-$6 of the three fibers is given in Figure \ref{fig:change_hygro} (Case I). The average FD SW fiber shrinkage and standard deviation of RH cycles 1$-$2 shown in Figure \ref{fig:strains_sw} (a) ($\bar{\epsilon}^{FD-}_{ll}$, $\bar{\epsilon}^{FD-}_{tt}$, and $\bar{\epsilon}^{FD-}_{lt}$) are added for reference. The fibers do not show a significant change in shrinkage. In short, soaking RD SW fibers for 12 hours does not induce a "transformation" to FD SW fibers. \\ \indent
To compliment the analysis, the twist of 25 randomly extracted fibers from each FD and RD, HW and SW handsheets was characterized. The SW fibers showed an average fiber twist of 0.60$\pm$0.37 and 0.08$\pm$0.04 rot/mm for FD and RD respectively, while the HW fibers showed a much lower difference in twist between FD and RD, i.e., 0.08$\pm$0.05 and 0.05$\pm$0.04 rot/mm respectively. After immersing each fiber in a droplet of water for 12 hours (while being completely unconstrained on a teflon sheet), the twist of the SW fibers increased to 0.80$\pm$0.39 and 0.38$\pm$0.18 rot/mm for FD and RD, respectively, and the HW fibers remained about constant at 0.09$\pm$0.0 0.06$\pm$0.04 rot/mm for FD and RD, respectively. The key finding is that the fiber twist of the RD SW fibers significantly increases after being soaked for a longer time, whereas the RD HW fibers do not. Hence, it is likely that the RD SW fibers require twisting in order to "transform" into a FD SW fiber and the RD HW fibers do not, which is consistent with the SW fibers' larger MFA compared to HW \citep{french2000effect, barnett2004cellulose, cown2004wood, donaldson2008microfibril}. Additionally, the increase in twist may indicate increased longitudinal stress, as this is converted into twisting due the MFA of the fiber. Hence, the minor constraint by the nylon threads may still have prevented the RD SW fibers from "transforming" into FD SW fibers after being soaked for 12 hours, as the fibers are only allowed to minimally rotate during the hygro-expansion experiments, and not the necessary twist increase of \textsuperscript{$\sim$}0.3 rot/mm as measured in the fiber twist study. \\ \indent
Therefore, the three SW RD fibers (from Case I) were again soaked in water for 12 hours, but this time without nylon threads making them completely free to twist. Note that all dry fibers demonstrated an increased fiber twist after being soaked without constraints, comparable to the numbers from the fiber twist study. These fibers were patterned, re-clamped by the threads, and subjected to two RH cycles from 30$-$90$-$30\% (cycles 7$-$8). $\bar{\epsilon}^-_{ll}$, $\bar{\epsilon}^-_{tt}$, and $\bar{\epsilon}^-_{lt}$ (with standard deviation) of the three fibers are given in Figure \ref{fig:change_hygro} (Case II). Note that none of the fibers show a significant change in shrinkage, and did not "transform" to FD SW fibers. Only a small increase in $\bar{\epsilon}_{lt}$ was found, which may logically be caused by the increased twist that the fibers now exhibit. It is possible that a mechanical activation (e.g. stirring) is required to structurally 'transform' the RD SW fibers into FD SW fibers. Therefore, the fiber structure is studied next. \\ \indent
To further explore the "transformation" from RD to FD fibers, the elastic stiffness of the fibers can also be studied. It is known that FD fibers yield a lower stiffness than RD fibers \citep{jentzen1964effect}. Hence to validate if fibers are hygroscopic and mechanically reversible, the elastic properties of RD fibers before and after wetting should be studied.

\subsection{The fiber's structural differences} 
In Figure \ref{fig:strains_1_2}, it is observed that the SW fibers exhibit a significantly lower $\bar{\epsilon}^-_{lt}$ than the HW fibers (for both RD and FD), even after soaking, considering that the helically-shaped micro-fibrils require to slide along each other to allow the hemi-cellulose to swell, and that the SW fiber's MFA is much larger than the HW fibers' \citep{french2000effect, barnett2004cellulose, cown2004wood, donaldson2008microfibril}. Additionally, $\bar{\epsilon}^-_{ll}$ of the SW fibers is only slightly higher than the HW fibers, where a much larger difference is expected according to the findings of \cite{meylan1972influence} and \cite{yamamoto2001model} for wood fibers. Therefore, it seems that the SW fibers tested here are somehow constrained in their movement. A plausible explanation could be that the lumen of the SW fibers are all (partially) collapsed. \\ \indent
For a collapsed lumen, the top and bottom cell wall are bonded to each other, in which the direction of the micro-fibrils of the bottom cell wall are crossed (under an angle of 2 times the MFA) with the top cell wall, restricting the fiber to shear and expand. This is in contrast to an open lumen case, which is more typical for wood fibers, thus explaining the difference between the present results and those of \cite{meylan1972influence} and \cite{yamamoto2001model}. A collapsed lumen can also explain the larger averaged value and variation in $\epsilon^-_{ll}$ and $\epsilon^-_{lt}$ for FD compared to RD SW fibers (see Figure \ref{fig:strains_sw} (a)). In the RD SW case, the lumen would logically more often be closed (compared to FD), restricting the $\epsilon^-_{ll}$ and $\epsilon^-_{lt}$ of the fiber, resulting in similar values. Whereas for the FD SW fibers the lumen is sometimes open, resulting in a larger scatter in $\epsilon^-_{ll}$ and $\epsilon^-_{lt}$. Such a hypothesis agrees well with the findings by \cite{he2003behavior}, who showed that larger drying pressures result in more frequent closed lumens, or sections of the lumen being closed. Considering the HW fibers, the frequency of a closed lumen is expected to be much less compared to the SW fibers, based on the higher fiber wall stiffness of the HW fibers having both a thicker cell wall and smaller fiber radius \citep{ilvessalo1995fiber, antes2015fiber}, making the structure less prone to collapsing. In contrast to the SW fibers which have a thin cell wall and large fiber radius \citep{lundqvist2002system}, which is consistent with the width to thickness ratio of, respectively, 3.2$\pm$1.5 and 5.9$\pm$3.1 for the HW and SW fibers tested here. Finally, a collapsed lumen of the SW fibers would also explain why the apparent MFA in Figure \ref{fig:strains_sw} (b) is approximately 0 degrees, because the top and bottom wall, which deform almost equally, constrain each other. As the lumen of the HW fibers is not collapsed, the top and bottom fiber walls are free to swell perpendicular to the micro-fibrils, resulting in a major strain direction, i.e. MFA, corresponding to literature values. \\ \indent
In summary, all these reasons make it quite plausible that the larger $\bar{\epsilon}^-_{lt}$ and minor difference in $\bar{\epsilon}^-_{ll}$ of the HW compared to SW fibers is because the lumen of the SW fibers is more often closed than the HW fibers due to the difference in the fiber's stiffness. To extend this theory towards the storage of dried-in strains, it is possible that closing the lumen during restrained drying locks the dried-in strain into the fibers, and the lumen needs to be (mechanically) opened in order to release the dried-in strain. This hypothesis may explain why the RD HW fibers are able to "transform" into FD HW fibers, whereas the RD SW fibers cannot. The details of the lumen of the fibers is therefore studied next. 

\subsubsection{Lumen analysis: experimental}
\begin{figure}[H]
	\centering
	\includegraphics[width=\textwidth]{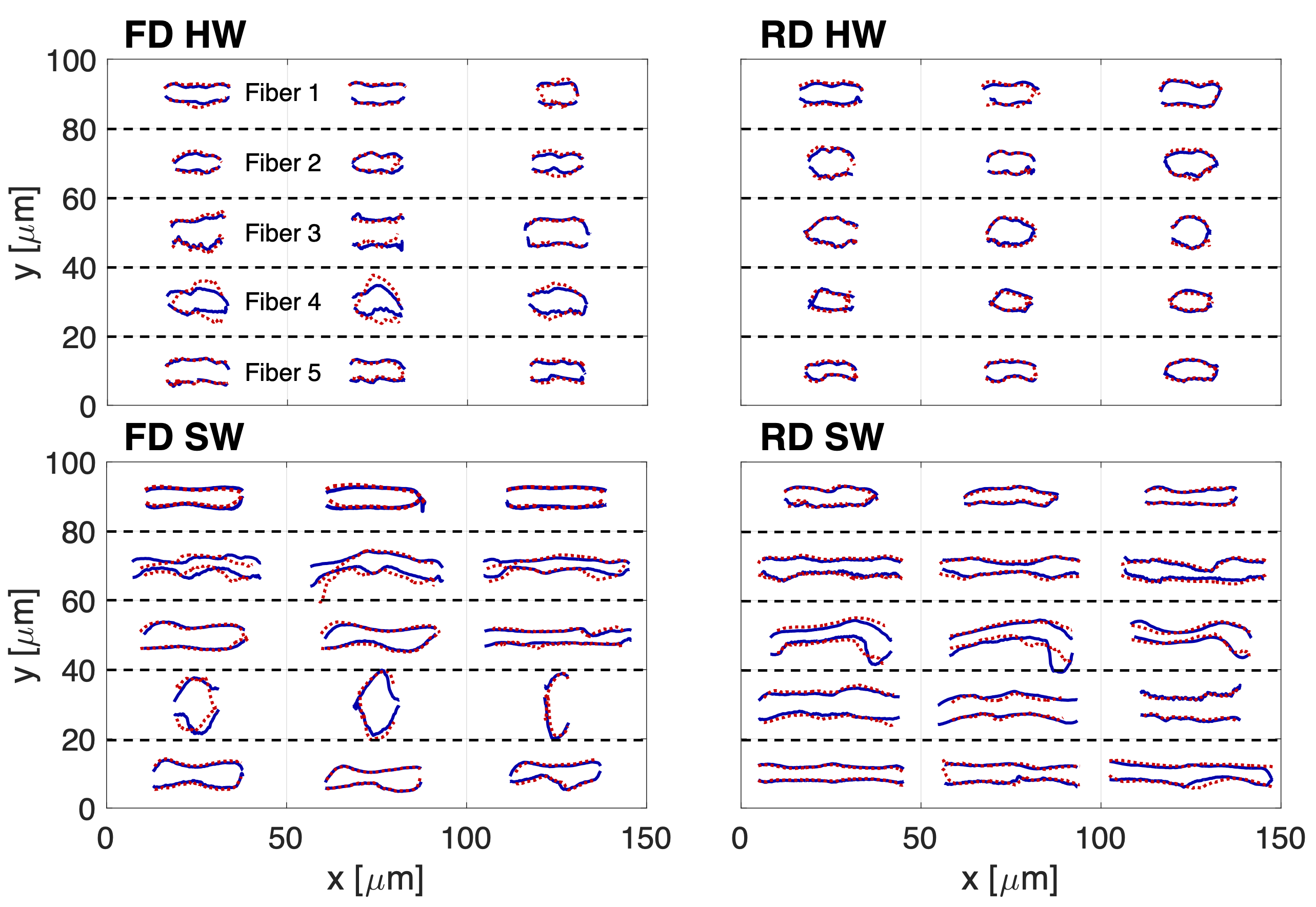}
	\caption{Change in cross-sectional shape of five FD and RD, HW and SW fibers characterized at three positions along the fiber length, before (blue) and after soaking (red) the fiber in water for 12 hours.}
	\label{fig:lumen}
\end{figure}

In order to study if the lumen of a fiber is (partially) open or closed, the change of the shape of the fiber is studied. For instance, the cross-sectional shape of an open lumen fiber is more likely to change when the fiber is analyzed before and after being soaked in water, while for a closed lumen, the shape is less likely to change. Hence, the shape of five FD and RD, HW and SW fibers (randomly extracted from the handsheets) at three different locations along the fiber length (at 25, 50 and 75\% of the fiber length) is analyzed before and after being soaked in water for 12 hours. This is done using the double sided imaging method proposed by \cite{Maraghechi2023mirror}, which was also used to characterize inter-fiber bonds in \citep{vonk2023bonds}. It involves capturing the front and back topography of the fiber at the target position, whereby the distance between the two topographies is known, hence enabling quasi-3D characterization of the fiber structure. Subsequently averaging the topography over the fiber length results in an average shape of the fiber in that area, as depicted before (blue) and after wetting (red) in Figure \ref{fig:lumen}. 
Three fiber sections are displayed per row for every fiber. The significant differences in fiber cross-sectional dimensions between the HW and SW fibers are directly visible. None of the HW fibers have a flat shape, both for FD and RD, confirming their open structure. The FD and RD SW fibers on the other hand have a much flatter structure, except for FD SW fiber 4, which most likely has an open structure. The RD SW fibers are closed both before, as well as, after soaking, visualized by their flat shape which remains unchanged. This explains why the hygro-expansivity of the RD SW fibers does not change after soaking, because the soaking treatment did not open the lumen. \\ \indent
The FD SW fibers are also mostly closed except for fiber 4, indicating that, on average, the FD SW fibers are less closed than the RD SW fibers. Moreover, all of the FD SW fibers show, to some extent, a shape change (more than RD SW fibers), which implies that the top and bottom can (partially) slide along each other. Perhaps the micro-fibrils decorating the top and bottom inside wall are entangled, keeping the lumen from completely opening, while still allowing some relative motion. The partial sliding ability of the top and bottom wall of the FD SW fibers would explain why the apparent MFA of the FD SW fibers, displayed in Figure \ref{fig:strains_sw} (b), is higher than that of RD SW fibers, but still significantly lower than the MFA values reported in literature. \\ \indent
This method enabled characterization of cross-sectional shape of the fiber before and after wetting. To analyze the shape of the lumen, a microtome can be used, similar to \cite{kappel2009novel}. This however does require embedding the fiber into a resin, and wetting will not be possible. 

\subsubsection{Lumen analysis: numerical}
\begin{figure}[]
	\centering
	\includegraphics[width=0.6\textwidth]{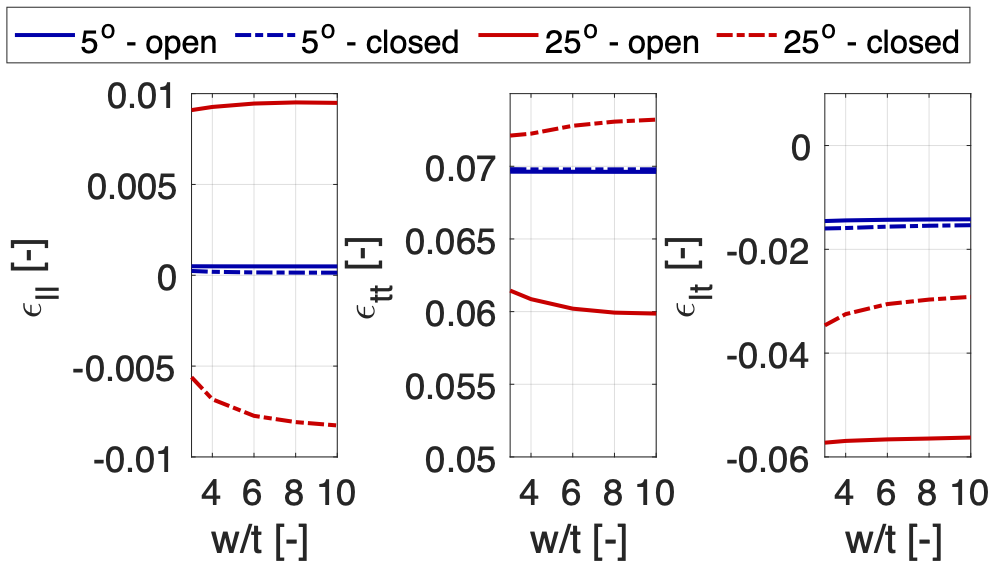}
	\caption{The longitudinal ($\epsilon_{ll}$), transverse ($\epsilon_{tt}$), and shear ($\epsilon_{lt}$) hygro-expansion of an open and closed fiber structure. An MFA of 5 and 25\textsuperscript{o} representing the average MFA of, respectively, HW and SW, and a width to thickness ratio ($w/t$) of 3-10 is studied. A schematic representation and details of the fiber model can be found in Subsection \ref{app:fiber_model}.}
	\label{fig:lumen_num}
\end{figure}

This above-described theory is validated using a numerical fiber model, of which a schematic representation along with the details are described in Subsection \ref{app:fiber_model}. Two extreme cases are considered, i.e. a fully open fiber and fully closed fiber. The effect of the fiber width to thickness ($w/t$) is studied for a range from 3 to 10, similar to the cross-sections in Figure \ref{fig:lumen}. Additionally, a MFA of 5 or 25\textsuperscript{o}, which are the average MFA values for the softwood and hardwood fibers tested here \citep{french2000effect, barnett2004cellulose, cown2004wood, donaldson2008microfibril}, are modeled. The $\epsilon_{ll}$, $\epsilon_{tt}$, and $\epsilon_{lt}$ at the fiber's top surface is plotted against the $w/t$ ratio, for closed and open lumen, and the MFA in Figure \ref{fig:lumen_num}. \\ \indent
First of all, the solutions tend to become constant for increasing $w/t$ ratios. As expected, for an open lumen, $\epsilon_{ll}$, and $\epsilon_{lt}$ increase for a larger MFA, similar to the wood fibers characterized by \cite{meylan1972influence} and \cite{yamamoto2001model}, whereas $\epsilon_{tt}$ logically decreases. Interestingly, for the fully closed lumen case, $\epsilon_{ll}$, and $\epsilon_{lt}$ decreases, which is most significant for the large MFA cases, where $\epsilon_{ll}$ even becomes negative. Hence, $\epsilon_{ll}$ and $\epsilon_{lt}$ decrease with increasing degree of bonding. Taking into account that the closed lumen model is of course an overestimation of the degree of internal fiber bonding (assumed to be driven by the same mechanisms as inter-fiber bonding, i.e. never perfectly bonded \citep{hirn2015comprehensive}), can explain why the SW fibers, in spite of their significantly larger MFA, exhibit a similar $\epsilon_{ll}$ as the HW fibers. The model also confirms the hypothesis that the lumen of the FD SW fibers is (partially) closed/open, while the lumen of the RD fibers is more often closed, directly explaining the larger scatter in $\epsilon^-_{ll}$ and $\epsilon^-_{lt}$ of the FD compared to the RD SW fibers displayed in Figure \ref{fig:strains_sw}. Finally, the result that a closed lumen can induce a negative $\epsilon_{ll}$ is confirmed by the SW pulp fiber tested in \citep{vonk2021full}, which exhibited a negative $\epsilon_{ll}$. Note that the proposed model does not predict the same hygro-expansion values as found in this work, as it lacks on some key aspects, i.e., non-continuous helical structure, no expansion along the micro-fibrils, no multiple cell wall layers., etc. \\ \indent
In summary, the above-conducted experimental and numerical analyses confirms that HW fibers, on average, have open lumen structures, and therefore exhibit (i) a relatively large longitudinal and shear hygro-expansion with a low MFA, similar to open wood fibers, and (ii) the "transformation" from a RD to a FD fiber can occur. The SW fibers, in which the lumen is partially closed for FD, and more often closed for RD, exhibit (i) a relatively low longitudinal and shear hygro-expansivity with a high MFA, in contrast to their wood fiber counterparts, and (ii) no "transformation" occurs from RD to FD.

\subsection{Fiber-to-sheet coupling}
\label{sec:sheet}
\begin{figure}[t!]
	\centering
	\includegraphics[width=0.5\textwidth]{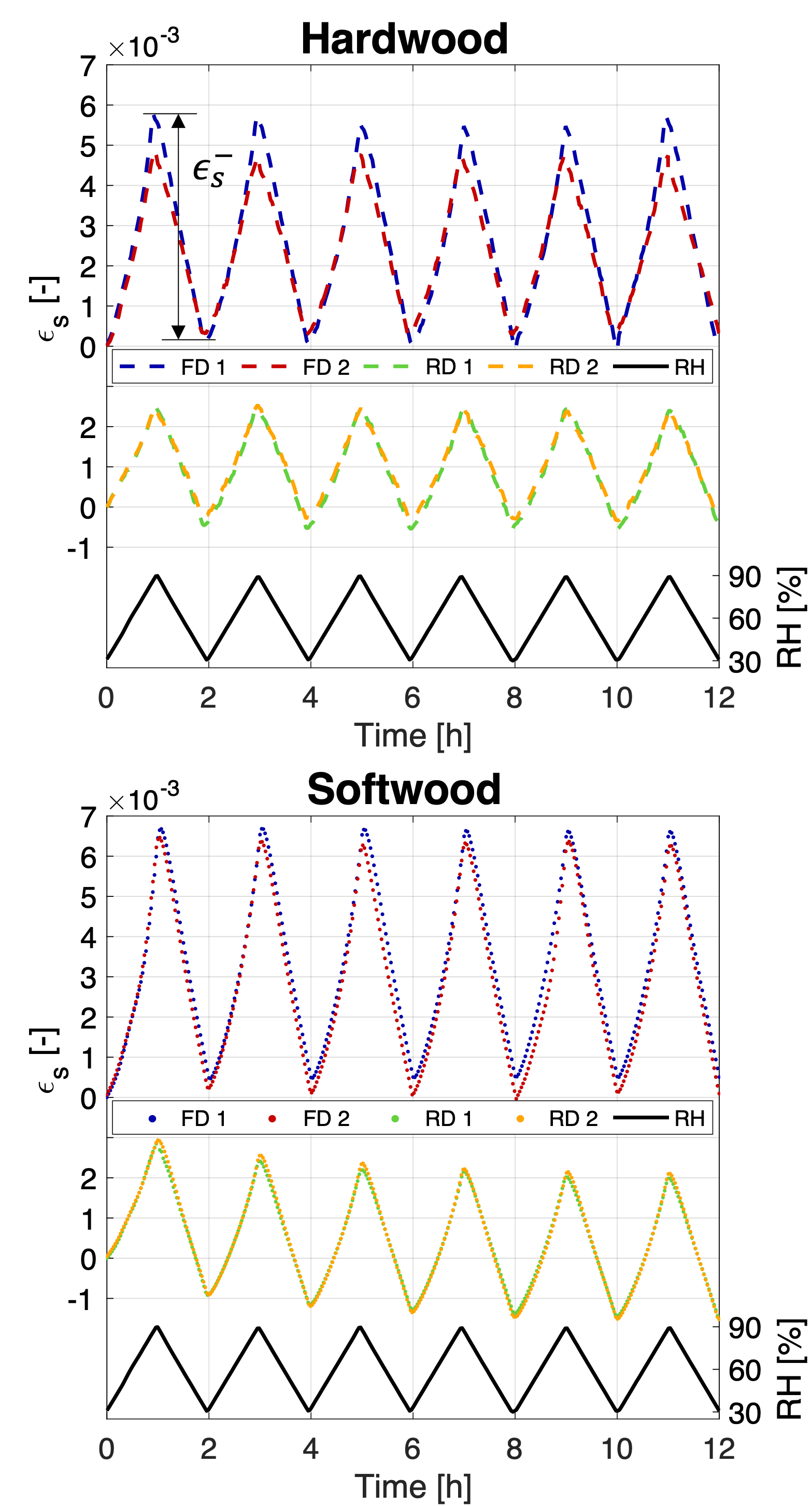}
	\caption{The isotropic hygro-expansion response ($\epsilon_s$) of two FD and RD SW and HW handsheets. The hygro-expansivity of (i) FD handsheets is significantly larger than then RD, and (ii) the SW handsheets is larger than the HW, both similar to literature \citep{larsson2008influence, urstoger2020microstructure, uesaka1992characterization, uesaka1997effects}. The shrinkage during each drying slope from 90 to 30\% RH, as annotated ($\epsilon^-_{s}$) in RH cycle 1 of HW FD1, is extracted and given in Table \ref{tab:sheet_strain}.}
	\label{fig:sheet_test}
\end{figure}

From modeling it is known that the longitudinal fiber hygro-expansion is dominant and fully contributes to the sheet hygro-expansion, while the transverse fiber strain contributes through the bonds \citep{brandberg2020role}. In order to experimentally validate this and separate the longitudinal and transverse fiber strain contributions to the sheet scale, the sheet-scale hygro-expansion evolution of the FD and RD, HW and SW handsheets, from which the fibers were extracted, is characterized and presented in Figure \ref{fig:sheet_test}. \\ \indent
The strains in both directions were equal for every RH cycle of every handsheet test, indicating the expected isotropic hygro-expansivity of the handsheets and the reliability of the measurement method. The isotropic sheet-scale hygro-expansion ($\epsilon_s$) response of the two RD and FD SW and HW handsheets is given in Figure \ref{fig:sheet_test}. The curves are in line with other sheet-scale hygro-expansion works, e.g., the RD handsheets show a clear release of irreversible dried-in strain after the wetting slope of the first RH cycles, whereas the FD handsheets do not \citep{larsson2008influence, uesaka1992characterization, niskanen1997dynamic}. Additionally, the RD SW handsheets display an ongoing release of dried-in strain characterized by every peak or valley per cycle is lower than the previous, in contrast to RD HW. \\ \indent
The average handsheet shrinkage ($\epsilon^-_{s}$) per handsheet, considering all six drying slopes as annotated for the first cycle ($\epsilon^-_{s}$), and the total average handsheet shrinkage ($\bar{\epsilon}^-_s$) considering both handsheets per pulp and drying procedure are given in Table \ref{tab:sheet_strain}. $\bar{\epsilon}^-_s$ is also added to the longitudinal fiber strain plot in Figure \ref{fig:strains_1_2}. $\bar{\epsilon}^-_s$ reveals that the HW and SW handsheets exhibit a factor 1.72$\pm$0.24 and 1.72$\pm$0.06 larger hygro-expansivity between FD and RD respectively, consistent with literature \citep{larsson2008influence, urstoger2020microstructure, uesaka1992characterization, uesaka1997effects}, and the SW handsheets exhibit a larger shrinkage, similar to the longitudinal and transverse fiber shrinkage presented in Figure \ref{fig:strains_1_2}. After adding $\bar{\epsilon}^-_s$ to the corresponding $\bar{\epsilon}^-_{ll}$ plot in Figure \ref{fig:strains_1_2}, $\bar{\epsilon}^-_{ll}$ of the fibers is close to the sheet-scale expansivity for every handsheet type, indicating that the contribution of the longitudinal fiber hygro-expansion to the sheet is indeed strong, while the transverse stain contribution is relatively weak. \\ \indent
From the fiber network model proposed by \cite{brandberg2020role}, a simple rule of mixture calculation can be used to upscale the fiber characteristics, given in Table \ref{tab:fiber_strain}, to find the sheet-scale hygro-expansivity, i.e., $1\cdot\epsilon_{ll}+0.0375\cdot\epsilon_{tt} = \epsilon_{s}$ ($(\beta_{s}(=0.0525)-\beta_{ll}(=0.03))/\beta_{ll}(=0.60) =0.0375$ in \citep{brandberg2020role}). For HW the predicted sheet-scale hygro-expansion is 0.0064$\pm$0.0005 and 0.0038$\pm$0.0009 for FD and RD, respectively, while for SW, it is 0.0068$\pm$0.0027 and 0.0051+0.0026 for FD and RD respectively. These predictions are slightly larger than the actual sheet-scale hygro-expansions given in Table \ref{tab:sheet_strain}. Moreover, the model predicts that the HW and SW handsheets exhibit a factor 1.67$\pm$0.52 and 1.33$\pm$1.20 larger hygro-expansivity between the FD and RD, respectively, whereas the experiment shows somewhat larger values, i.e. 1.72$\pm$0.24 and 1.72$\pm$0.06 respectively. The small differences may be attributed to the difference in paper structure between the model and experiment, i.e., fiber coverage, grammage, fiber geometry, etc., all affecting the model's outcome. \\ \indent
The fact that (i) the longitudinal fiber hygro-expansion is indeed dominant at the sheet scale, (ii) the longitudinal fiber hygro-expansion is strongly reduced by the RD process, and (iii) the dried-in strain stored inside the SW fibers is difficult to release is actually beneficial for digital printing purposes. Indeed any decrease of the hygro-expansivity of the fibers directly reduces the severity of the unwanted out-of-plane deformations such as cockling, fluting, waviness, and curl.

\section{Conclusion}
The significantly larger hygro-expansivity of freely compared to restrained dried paper sheets has been frequently studied in the literature. Various theories or hypotheses have been forwarded to explain these differences, including the geometry of the inter-fiber bonds and the structural fiber changes due to the drying procedure. The latter theory is studied in this work by testing the hygro-expansion of fibers extracted from freely and restrained dried handsheets. \\ \indent
To this end, restrained dried and freely dried handsheets were produced from either hardwood or softwood fibers. The hygro-expansion of the handsheets was obtained using a novel sheet-scale hygro-expansion method. Single fibers were extracted from the remainder of the handsheets which were tested using a recently-developed single fiber hygro-expansivity method which captures the transient full-field (longitudinal, transverse and shear) hygro-expansion during relative humidity (RH) changes. Each fiber was subjected to (i) two 30$-$90$-$30\% RH cycles, then (ii) a wetting cycle in which the RH is increased to 95\% for a longer time, while the fiber is cooled down, initiating hydro-expansion and maximizing the fiber's moisture content, and finally (iii) two 30$-$90$-$30\% RH cycles. This strategy is chosen to see if a restrained dried fiber is able to "transform" into a freely dried fiber. \\ \indent
The freely dried handsheets exhibit a significantly larger hygro-expansivity than the restrained dried. Furthermore, it was found that the restrained drying procedure mainly lowers the fibers' longitudinal hygro-expansion, which is dominant at the sheet-scale hygro-expansion, hence explaining the sheet-scale hygro-expansion differences. Regarding the hardwood fibers:
\begin{itemize}
	\item computation of the principal strains revealed that the major-minor strain angle (perpendicular to the micro-fibrils) and the minor strain (along the micro-fibrils) is lower for restrained dried, hence the lower longitudinal hygro-expansion,
	\item the lower minor strain of restrained dried fibers compared to the freely dried fibers confirm an older theory from the literature,
	\item all fibers were able to "transform" from restrained to freely dried.
\end{itemize} 
Regarding the softwood fibers: 
\begin{itemize}
	\item the principal strain computation did not show any change in the major-minor strain angle between the freely and restrained dried fibers,
	\item the fibers were not able to "transform" from a restrained to a freely dried fiber, even after immersing the fiber in water in an unconstrained condition for long periods. 
\end{itemize}
To study the driving mechanisms behind these differences between the hardwood and softwood fibers, the fibers' cross-sectional shape before and after wetting was investigated experimentally and numerically to elucidate the influence of the lumen. On average, the lumen of the hardwood fibers revealed to be open, while the softwood fibers are closed. For softwood, the micro-fibrils of the top and bottom wall are crossed and bonded, which (i) restrict free hygro-expansion, resulting in different major-minor strain angles, and (ii) restricts the release of dried-in strain, which requires the lumen to be open in order to be released. Finally, the strength of testing single fibers from handsheets enables a direct comparison between the fiber and sheet properties, and it was found that the transverse fiber hygro-expansion contribution to the sheet scale is relatively low.

\section{Acknowledgments}
The authors would like to acknowledge Marc van Maris of Eindhoven University of Technology for lab support and technical discussions. Also, the authors would like to acknowledge Louis Saes and Thomas Anijs of Canon Production Printing for extensive technical discussions and suggestions.
\section{Author contributions}
NV: Conceptualization, Methodology, Software, Validation, Investigation, Writing – original draft, Visualization. RP: Methodology, Supervision, Funding acquisition. MG: Methodology, Resources, Writing – review \& editing, Supervision, Funding acquisition. JH: Conceptualization, Methodology, Validation, Resources, Writing – review \& editing, Supervision, Funding acquisition.
\section{Funding}
This work is part of an Industrial Partnership Programme (i43-FIP) of the Foundation for Fundamental Research on Matter (FOM), which is part of the Netherlands Organization for Scientific Research (NWO). This research programme is co-financed by Canon Production Printing, University of Twente, Eindhoven University of Technology.
\section{Data availability}
No data is provided with this manuscript
\section{Competing interests}
The authors declare that they have no known competing financial interests or personal relationships that could have appeared to influence the work reported in this work.
\section{Consent of publication}
The authors hereby consent to publication of the present research work in this journal, if selected for publication.
\section{Ethics approval}
This article does not contain any studies with human participants or animals performed by any of the authors.

\appendix \newpage
\setcounter{figure}{0} 
\setcounter{table}{0} 
\setcounter{page}{1}

\section{Supplementary material}
\subsection{Fiber hygro-expansion magnitudes}
\label{sec:app1}
\begin{figure}[b!]
	\centering
	\includegraphics[width=\textwidth]{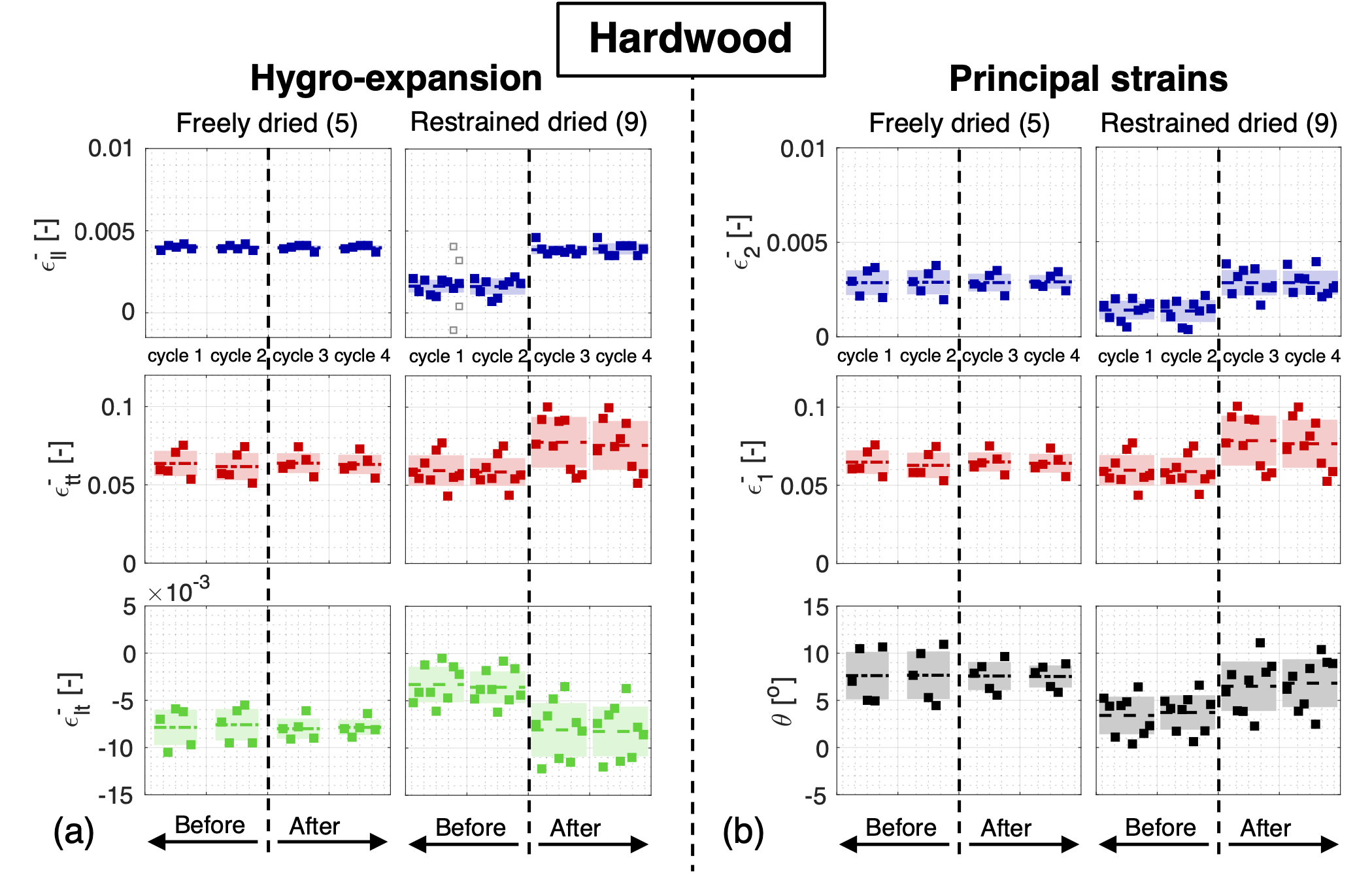}
	\caption{(a) The longitudinal ($\epsilon^-_{ll}$) and transverse shrinkage ($\epsilon^-_{tt}$), and shear strain change ($\epsilon^-_{lt}$) during the drying slopes from 90 to 30\% RH per cycle of each HW fiber separately (five and nine fibers extracted from respectively, the FD and RD HW handsheets), as annotated for the FD HW fiber curve in Figure \ref{fig:trend_fiber}. The fiber numbering is the same as the separate dried-in strain release values given in Figure \ref{fig:DI}. The gray markers in RD $\epsilon^-_{ll}$ plot represent the shrinkage during the drying (top marker) and wetting (bottom marker) of the fibers that exhibited severe ongoing shrinkage during both wetting and drying of cycle 1, for which the blue marker represents the average value considered the analysis. (b) The principal strains ($\epsilon_1$ and $\epsilon_2$) and major-minor strain angle ($\theta$) are computed from the full-field hygro-expansion, where $\theta$ approximates the MFA.}
	\label{fig:app_hw}
\end{figure}

\begin{figure}[b!]
	\centering
	\includegraphics[width=\textwidth]{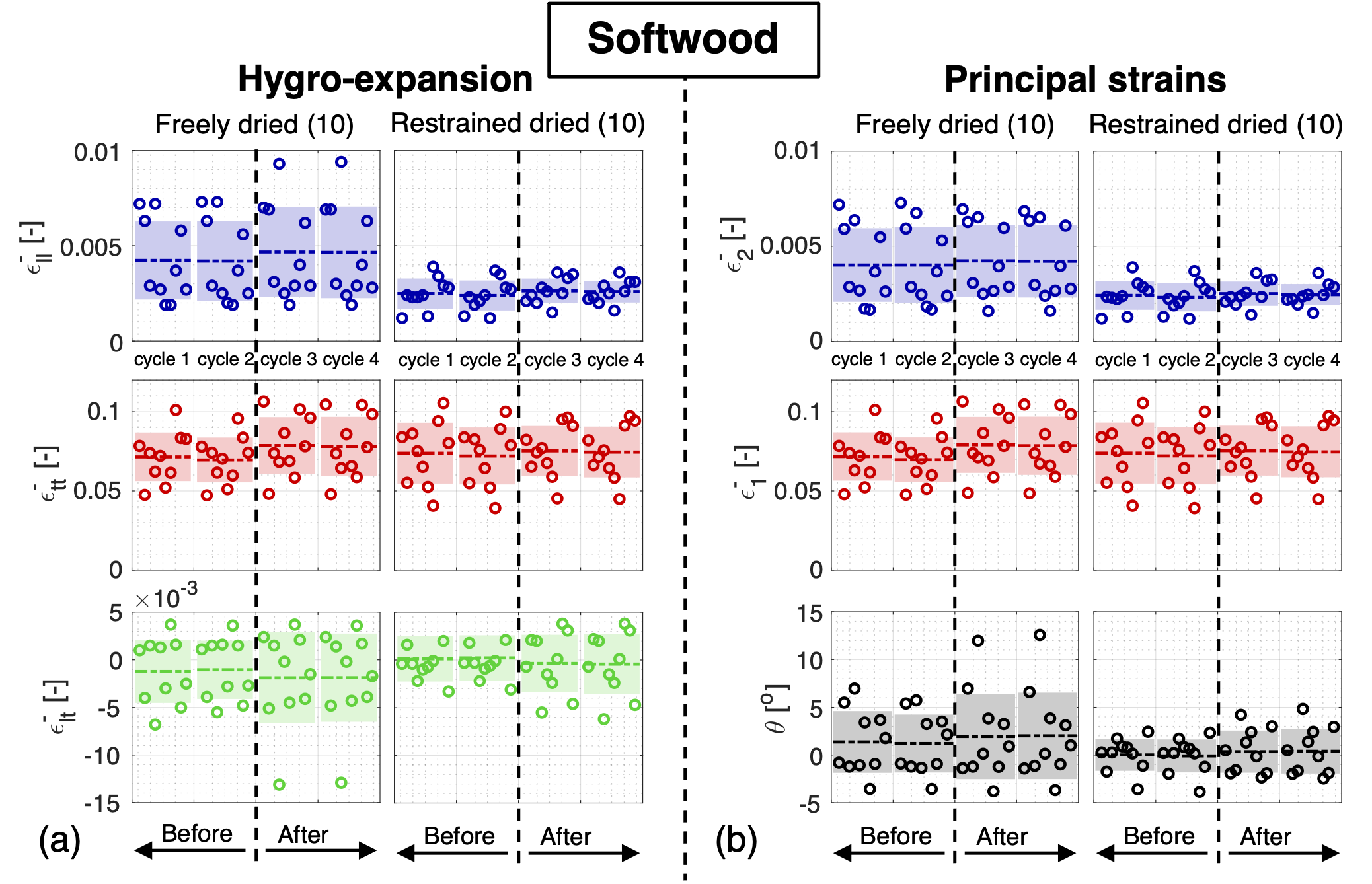}
	\caption{(a) The longitudinal ($\epsilon^-_{ll}$) and transverse shrinkage ($\epsilon^-_{tt}$), and shear strain change ($\epsilon^-_{lt}$) during the drying slopes from 90 to 30\% RH per cycle of each SW fiber separately (ten fibers extracted from each FD and RD SW handsheet), as annotated for the FD HW fiber curve in Figure \ref{fig:trend_fiber}. The fiber numbering is the same as the separate dried-in strain release values given in Figure \ref{fig:DI}. (b) The principal strains ($\epsilon_1$ and $\epsilon_2$) and major-minor strain angle ($\theta$) are computed from the full-field hygro-expansion, where $\theta$ approximates the MFA.}
	\label{fig:app_sw}
\end{figure}

\setcounter{table}{0} 
The longitudinal ($\epsilon^-_{ll}$) and transverse shrinkage ($\epsilon^-_{tt}$), and shear strain change ($\epsilon^-_{lt}$) during the drying slopes from 90 to 30\% RH per cycle of each resulting HW and SW fiber are separately given in Figure \ref{fig:app_hw} and \ref{fig:app_sw} respectively. The marker indicates the fiber number, i.e., the first marker corresponds to fiber 1, and so on, consistent with the marker numbering of release of dried-in given in Figure \ref{fig:DI}. \\ \indent
After testing, three RD HW fibers showed a distinct behavior from the nine other fibers, i.e. the shrinkage was equal to the value obtained from the FD HW fibers, indicating that these fibers did not carry any load during the RD procedure, and consequently dried in a freely manner. As the uncertainty around these data points would compromise the analysis, these fibers are excluded from all here-presented analyses, still leaving nine reliable RD HW fibers. Note that none of the RD SW fibers showed this phenomenon, which can be explained by the fiber length difference (HW fibers: \textsuperscript{$\sim$}1 mm versus SW fibers: \textsuperscript{$\sim$}3 mm), making it unlikely that any SW fiber remained completely free inside the fibrous structure. Furthermore, two RD HW fibers revealed significant ongoing shrinkage during the first RH cycle, entailing a relatively large $\epsilon^-_{ll}$ when only the drying slope is considered. One fiber even showed shrinkage during the wetting slope of the first wetting cycle.
Therefore, the average of the wetting and drying slope (shown by the gray markers in Figure \ref{fig:app_hw}(a)) is considered as the fiber's $\epsilon^-_{ll}$. It is presumed that these fibers may have been subjected to an extra large tensile force during paper formation, which subsequently triggered relaxation during wetting. Finally, three RD HW fibers did not exhibit the "transformation" after the wetting cycle. The fibers did, however, "transform" after keeping them in 100\% RH for 12 hours. The resulting average shrinkage ($\bar{\epsilon}^-_{ll}$, $\bar{\epsilon}^-_{tt}$, and $\bar{\epsilon}^-_{lt}$) during the drying slopes of RH cycles 1$-$2, before the wetting period, of all fibers is given Table \ref{tab:fiber_strain}.

\begin{table}[H]
	\centering
	\caption{Average longitudinal and transverse shrinkage, and shear strain change, i.e. $\bar{\epsilon}^-_{ll}$, $\bar{\epsilon}^-_{tt}$, and $\bar{\epsilon}^-_{lt}$, with standard deviation of all FD and RD, HW and SW fibers considering the first two drying slopes from 90 to 30\% RH before the wetting period.}
	\label{tab:fiber_strain}
	\resizebox{0.5\textwidth}{!}{
		\begin{tabular}{@{}cccc@{}}
			\hline
			\textbf{FD} & $\bar{\epsilon}^-_{ll}$     [-]        & $\bar{\epsilon}^-_{tt}$ [-]   & $\bar{\epsilon}^-_{lt}$    [-]     \\ \hline
			HW & 0.0040$\pm$0.0002 & 0.0627$\pm$0.0089 & -0.0077$\pm$0.0019 \\
			SW & 0.0042$\pm$0.0021 & 0.0705$\pm$0.0151& -0.0011$\pm$0.0033  \\ \hline \hline
			\textbf{RD} & $\bar{\epsilon}^-_{ll}$     [-]        & $\bar{\epsilon}^-_{tt}$ [-]   & $\bar{\epsilon}^-_{lt}$    [-]      \\ \hline
			HW & 0.0016$\pm$0.0005 & 0.0589$\pm$0.0096 & -0.0034$\pm$0.0018\\
			SW & 0.0024$\pm$0.0019 & 0.0729$\pm$0.0191 & 0.0002$\pm$0.0024 \\ \hline
	\end{tabular}}
\end{table}

\subsection{Determination of the micro-fibril angle}
\label{app:mfa}
\begin{figure}[b!]
	\centering
	\includegraphics[width=0.5\textwidth]{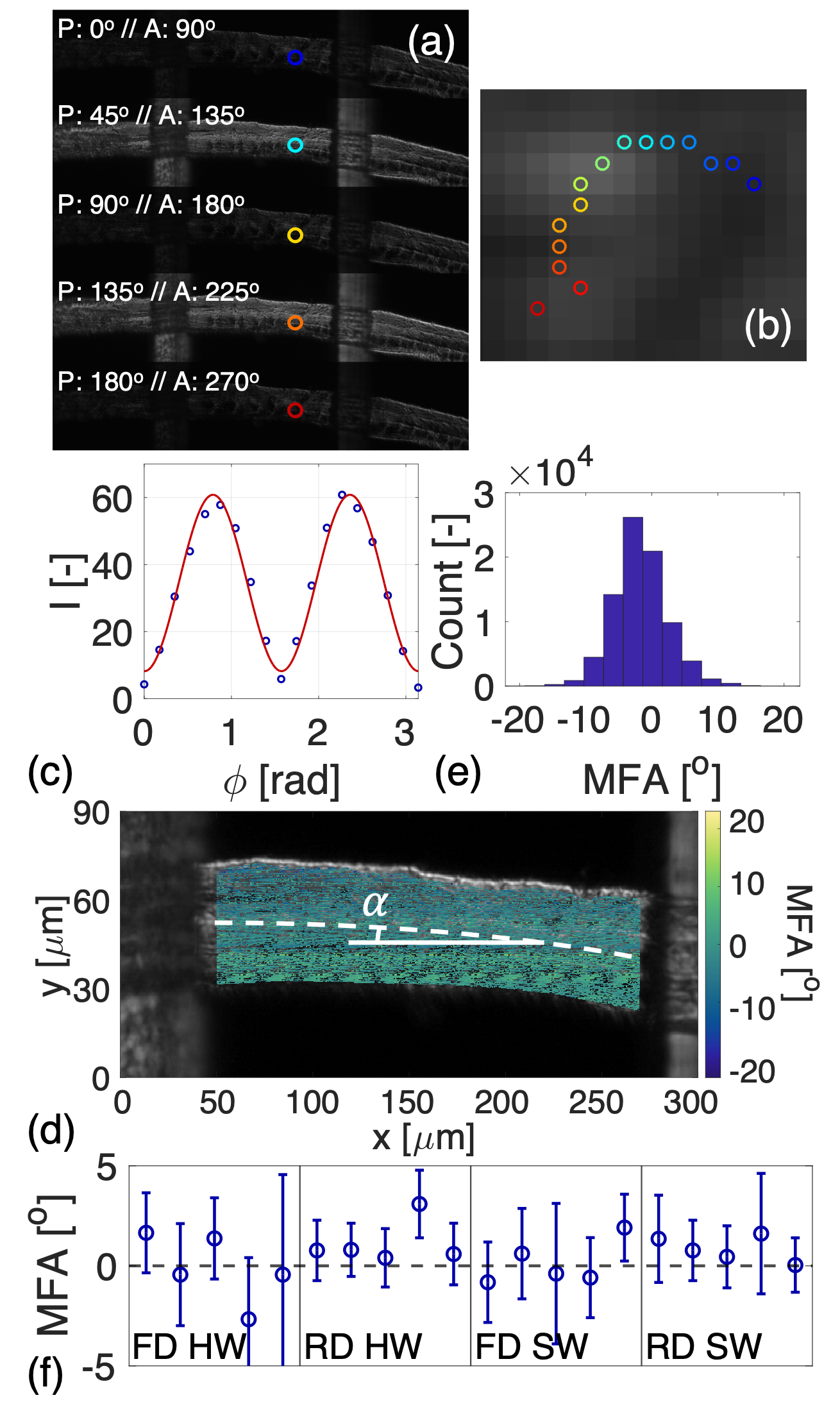}
	\caption{Improved MFA measurement method with (a) five images at different polarizer and analyzer rotations, (b) the image shift due to misalignment of the polarizers, (c) the intensity profile of a pixel during rotation of 180$^o$ of the crossed polarizers including the fit (Equation \ref{eq:mfa}), (d) the MFA for every pixel (R\textsuperscript{2} $>$ 0.9) and the fiber angle with respect to the horizontal axis ($\alpha$), (e) an histogram showing the scatter of the MFA over the full fiber around -3\textsuperscript{o} , and (f) the MFA determination of 5 FD and RD, HW and SW fibers scattered around zero, indicating that the method measures the average of the front and back cell wall which are crossed, highlighting the unreliability of the MFA measurement method.}
	\label{fig:mfa}
\end{figure}

The method is based on the premise that the light intensity under polarization microscopy, which measures the birefringence of a material, should theoretically be zero when the in-plane orientation of linearly polarized light is aligned with the (average) MFA. Therefore, the method involves placing a polarizer (initially horizontal) and crossed analyzer (initially vertical), respectively, before and after the specimen in the light path of an optical microscope, and simultaneously rotating both polarizer and analyzer around the optical axis (up to 180 degrees), while an intensity image is obtained every 10\textsuperscript{o}, see Figure \ref{fig:mfa} (a). By doing so, a light intensity versus rotation profile is measured for every pixel of the fiber from which the MFA can be deducted by fitting a squared sine wave with phase shift to this profile. Yet, during the rotation of the polarizers, the projected image exhibits lateral shifts in the order of a few tens of pixels due to misalignment of the polarizers, deteriorating the reliability of the method. This problem is remedied here by correlating the images using GDIC with a 0\textsuperscript{th} order polynomial to correct for this rigid body motion and align the images with better than 0.1 pixel accuracy, see Figure \ref{fig:mfa} (b), resulting in a reliable (aligned) intensity profile per pixel, as shown for a single pixel in Figure \ref{fig:mfa} (c). The intensity profile of each pixel is subsequently fitted using the squared sine function: 
\begin{equation}
	I = A\text{sin}^2(2(\phi-\alpha-\text{MFA}))+B,
	\label{eq:mfa}
\end{equation}
in which $I$ is the intensity, $A$ the amplitude, $\phi$ the period ($0<\phi<\pi$), $\alpha$ the angle of fiber with respect to the horizontal axis for each pixel, as depicted in Figure \ref{fig:mfa} (d), and $B$ the vertical intensity shift. An example of such a fit is shown in Figure \ref{fig:mfa} (c). As a result, the MFA is obtained for every pixel and plotted for a SW fiber in Figure \ref{fig:mfa} (d) (R\textsuperscript{2} $>$ 0.9). The histogram displayed in Figure \ref{fig:mfa} (e) shows a distribution of the MFA of each pixel around a mean value of \textsuperscript{$\sim$}-3\textsuperscript{o}, of which the absolute value (fiber may be rotated) is outside the expected range of 8$-$39\textsuperscript{o} \citep{barnett2004cellulose, cown2004wood} reported in literature. Analysis of 5 FD and RD, HW and SW fibers resulted in average MFA values scattered around 0\textsuperscript{o} for every fiber type as given in Figure \ref{fig:mfa} (f). At first glance, these results may seem flawed. However, when the phase shift in the light intensity versus rotation profile indeed corresponds to the averaged MFA, as should be the case from theory, then a measurement where the light beam passes through a fiber should give an average MFA of zero, because the MFA of the front and back fiber wall being precisely opposite. Hence, it is concluded that this transmission polarization microscopy-based MFA measurement method, despite improving it with the GDIC-based image alignment, is fundamentally unable to measure the MFA of fibers with opposite MFA in the front and back fiber wall, which is always the case for helical fibers. This method may however work when a single fiber wall is characterized, which may be obtained by carefully milling the fiber from one side using, e.g., a broad ion gun, which has not been explored. Moreover, due to limitations on the research project, no other direct MFA measurement methods could be assessed. 

\subsection{Numerical fiber model}
\label{app:fiber_model}
\begin{figure}[H]
	\centering
	\includegraphics[width=0.75\textwidth]{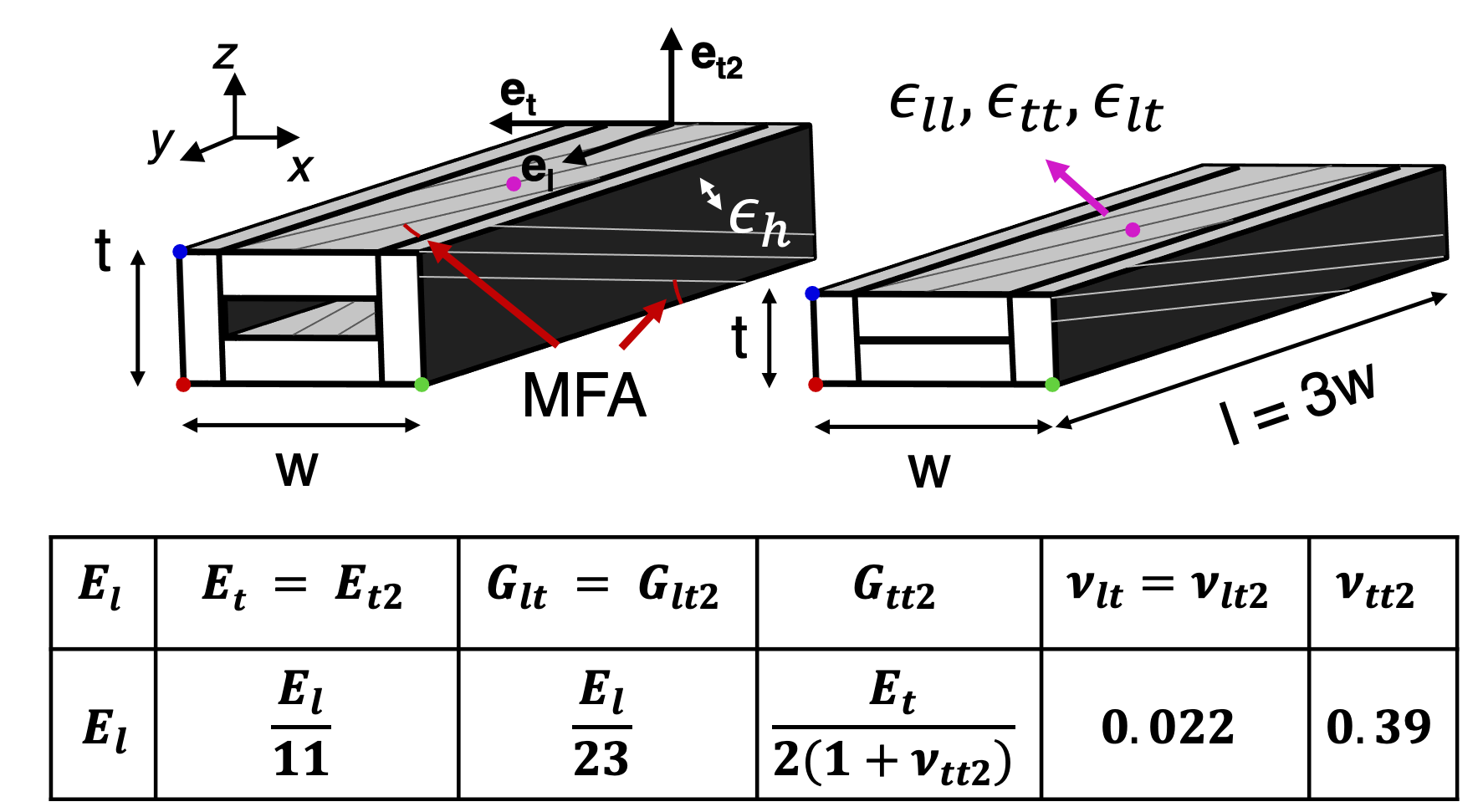}
	\caption{Schematic representation of the numerical fiber model consisting of four oriented rectangular bodies to reflect the helical fibril structure of a paper fiber. An open and closed lumen case is considered. The material properties are given in the table, which are the same as used by \cite{magnusson2013numerical}. The deformation is driven by the strain in the direction perpendicular to the micro-fibrils ($\epsilon_h$). Boundary conditions: red node: displacement is constrained in all three direction, green node: in $y$ and $z$-direction, and blue node: in $y$-direction. The strain values are obtained at the center of the top fiber surface (purple node), and plotted in Figure \ref{fig:lumen_num}.}
	\label{fig:fiber_model}
\end{figure}

A numerical fiber model is realized (in \textit{MSC. Marc}) to study the effect of (i) an open or closed lumen, (ii) the fiber cross-sectional dimensions, and (iii) the MFA on the hygro-expansivity of paper fibers, which is schematically displayed in Figure \ref{fig:fiber_model}. The model consists of four rectangular bodies which each have their own orientation, representing the helical micro-fibril structure of the fiber wall. For the open lumen case, the bodies (modeled as contact bodies) are bonded to each other such that an open tube-like structure is realized. The height of the lumen is chosen to be four times the fiber wall thickness, which remains constant for every simulation. For the closed lumen case, the bodies are perfectly bonded to each other. \\ \indent
The deformation of the simulation is driven by the hygroscopic strain in between the micro-fibrils ($\epsilon_h$) which is chosen at 7\% in both directions perpendicular to the fibril orientation, similar to the $\bar{\epsilon}^-_{2}$ values found in Figures \ref{fig:app_hw} and \ref{fig:app_sw}. The hygro-expansion along the micro-fibrils is set to zero. The constrained boundary nodes are presented by the colored nodes in Figure \ref{fig:fiber_model}, i.e. the displacement of the red node is constrained in all three directions, the green node in $y$ and $z$-direction, and the blue node only in $y$-direction. Both models were simulated using quadratic cubic elements, and the material properties are given in the table in Figure \ref{fig:fiber_model}, which are the same as used by \cite{magnusson2013numerical}.\\ \indent
The width to thickness ratio ($w/t$) is varied from 3 to 10, because, as Figure \ref{fig:lumen} shows, the SW have a significantly larger $w/t$ ratio than the HW fibers, which may also affect the fiber hygro-expansion. Note that the thickness is kept constant while the width is changed. The length of the fiber is equal to three times the width of the fiber for both open and closed lumen. Furthermore, two MFA cases are modeled, i.e. 5 and 25\textsuperscript{o}, representing the average MFA of, respectively, HW and SW \citep{french2000effect, barnett2004cellulose, cown2004wood, donaldson2008microfibril}. The longitudinal, transverse and shear hygro-expansion curves along the central part (25 to 75\%) of the fiber remained almost constant, implying that the fiber length used in the model is sufficient to ignore boundary effects. The surface hygro-expansion is extracted from the center of the top fiber surface, see the purple node in Figure \ref{fig:fiber_model}, and is plotted for every $w/t$ ratio, closed or open lumen, and MFA in Figure \ref{fig:lumen_num}.

\subsection{Handsheet hygro-expansion magnitudes}
\label{app:sheet_exp}

\begin{table}[H]
	\centering
	\caption{The average shrinkage per handsheet ($\epsilon^-_s$) considering six drying slopes from 90 to 30\% RH as depicted for the first cycle of HW FD1 in Figure \ref{fig:sheet_test}, and the total average shrinkage ($\bar{\epsilon}^-_s$) of the two handsheets per pulp and drying procedure combined. $\bar{\epsilon}^-_s$ of each handsheet type is added to the respective $\epsilon_{ll}^-$ plot in Figure \ref{fig:strains_1_2}.}
	\label{tab:sheet_strain}
	\resizebox{0.35\textwidth}{!}{
		\begin{tabular}{@{}cc|c@{}}
			\hline
			\textbf{HW }& $\epsilon^-_{s}$ [-] & $\bar{\epsilon}^-_{s}$ [-]  \\ \hline
			FD1   & 0.0056$\pm$0.0001 & \multirow{2}{*}{0.0050$\pm$0.0006}\\
			FD2  & 0.0045$\pm$0.0001 & \\ \hline
			RD1  & 0.0030$\pm$0.0000 & \multirow{2}{*}{0.0029$\pm$0.0001}\\
			RD2  & 0.0029$\pm$0.0001 &\\ \hline \hline
			\textbf{SW} & $\epsilon^-_{s}$ [-] & $\bar{\epsilon}^-_{s}$ [-]  \\ \hline
			FD1  & 0.0061$\pm$0.0001 & \multirow{2}{*}{0.0062$\pm$0.0001}\\
			FD2  & 0.0062$\pm$0.0001 & \\ \hline
			RD1  & 0.0035$\pm$0.0001 & \multirow{2}{*}{0.0036$\pm$0.0001}\\
			RD2  & 0.0037$\pm$0.0001 & \\ \hline
	\end{tabular}}
\end{table}

\end{document}